\documentclass[aps,prd,superscriptaddress,showpacs,twocolumn,nofootinbib]{revtex4-1}
\usepackage{graphicx}
\usepackage[caption=false]{subfig}
\usepackage{epstopdf}
\usepackage{amsmath}
\usepackage{amsfonts} 
\usepackage{amssymb}
\usepackage{latexsym}
\usepackage{hyperref}
\usepackage[english]{babel}
\usepackage[utf8]{inputenc}
\usepackage[dvipsnames]{xcolor}
\usepackage{slashed}
\usepackage{feynmp}
\usepackage{bm}
\usepackage{bbold}
\usepackage{eufrak}
\usepackage{tabu}
\usepackage{tikz}

\usepackage{bigstrut}

\begin{document}

\title{Shining massive light through a wall}

\author{P. C. Malta}\email{pedrocmalta@gmail.com}
\affiliation{R. Antonio Vieira 23, 22010-100, Rio de Janeiro, Brazil}



\begin{abstract}

A massive photon possesses a longitudinal polarization mode absent in its massless counterpart. Transverse and longitudinal modes follow different dispersion relations, the latter being much less attenuated than the former when passing through a conductor, suggesting the possibility of isolating longitudinal modes by shining intense light on a conducting wall. We calculate the transmission rates for normal incidence upon a semi-infinite medium and passage through a slab. For the second case we compare the expected photon fluxes with those measurable in current and future light-shining-through-a-wall experiments. Using a 1-MW microwave source as envisaged by the STAX project, a sensitivity at the level of $m_\gamma < 9.6 \times 10^{-11} \, {\rm eV/c^2}$ could be reached after a run time of an year, with a potential improvement by a factor of $\sim 10^4$ if radio waves of similar power are used.

\end{abstract}

\maketitle


\section{Introduction}  \label{sec_intro}
\indent

The photon is the paradigm for a massless particle. In fact, the invariance of electromagnetism under local gauge transformations requires $m_\gamma = 0$, though the photon may indeed acquire an effective mass in certain circumstances, such as in the vicinities of collapsed stars~\cite{Skobelev, Cover}, or in superconductors~\cite{supercond}. The fundamental masslessness of the photon is a theoretically pleasing hypothesis and is usually taken for granted, but it must be thoroughly tested in the laboratory and via astrophysical observations.

The most recent upper limits are listed by the Particle Data Group~\cite{PDG}, where the strongest limit quoted is $m_\gamma \leq 10^{-18} \, {\rm eV/c^2}$, coming from the analysis of solar-wind data collected by the Voyager missions~\cite{MHD}. Other strong upper bounds were extracted using fast radio bursts~\cite{Bonetti, Bentum, Wang}, solar-wind data at Earth's orbit~\cite{Retino}, Jovian magnetic-field measurements~\cite{Davis} and tests of Coulomb's law~\cite{Williams}. For reviews, see refs.~\cite{Tu, Nieto, Okun, Goldhaber,Goldhaber2}. Terrestrial phenomena may also be used to establish upper bounds. Fischbach found $m_\gamma \leq 8 \times 10^{-16} \, {\rm eV/c^2}$ using geomagnetic data~\cite{Fischbach} and Kroll studied Schumann resonances to obtain $m_\gamma \leq 2.4 \times 10^{-13} \, {\rm eV/c^2}$~\cite{Kroll1, Kroll2}; this limit has been recently updated to $m_\gamma \leq 2.5 \times 10^{-14} \, {\rm eV/c^2}$~\cite{Malta}.

Goldhaber and Nieto emphasized that measuring the photon mass is only possible if the measuring apparatus or dimension of the system under consideration are very large or the measurement is very precise~\cite{Goldhaber, Goldhaber2,Nieto}. In particular, the photon mass only appears squared and effects are $\mathcal{O}(m_\gamma^2 d^2)$, where $d$ is the typical size of the system under consideration. The upper limits quoted above support this statement and astrophysical limits are tighter than terrestrial ones. It is nonetheless important to search for novel -- and exquisitely precise -- laboratory experiments that could allow us to reach the currently best terrestrial (and perhaps astrophysical) limits with the added benefit of precisely controlling the light source, the environment and the detector system.

One of the main features of a massive photon is the existence of a third, longitudinal polarization mode. Interestingly enough, this mode has a different dispersion relation than the other two transverse modes when propagating in a medium other than vacuum. In a conductor, the wave number of longitudinal modes is much more weakly influenced by the finite conductivity than that of transverse modes. This means that the reflection and transmission properties will be significantly different depending on the polarization of the incident wave.

In this work we explicitly calculate the reflection and transmission rates of de Broglie-Proca waves when crossing a single interface into a semi-infinite medium. More importantly, we consider the experimentally interesting case of propagation through a conducting slab. Given the disparity between the in-medium behaviors of longitudinal and transverse modes, we expect to find different rates depending on the polarization mode in question. In particular, we will show that good conductors are effectively transparent to longitudinal waves, whereas transverse waves behave very similarly to their massless counterparts, being strongly reflected at the interface and experiencing exponential attenuation inside the conductor. This suggests that, in a situation where all modes are present, a conducting slab would strongly reflect the transverse components while letting the longitudinal modes pass unattenuated, thus acting as a filter. Placing a sensitive enough photon detector behind the slab would then allow us to detect the few longitudinal photons that successfully go through, thereby effectively measuring massive light shining through a wall.

The concept of a light-shining-through-a-wall (LSW) experiment was first proposed in the 1980s to search for hypothetical particles coupled to the electromagnetic field, such as axions~\cite{LSW_orig, LSW_orig2,LSW_orig3}. In these experiments a light source -- typically a high-power laser -- shines onto an optical barrier, the wall. Turning on a magnetic field in the region before the wall allows some of the photons to be converted into these hypothetical particles. Given their weak coupling to matter, they transverse the barrier unscathed. Placing an equally strong magnetic field behind the wall stimulates part of these particles to reconvert into photons with a small, but finite probability~\cite{Adler, Redondo}.

Several implementations of LSW experiments have been pursued, among them OSQAR~\cite{OSQAR, OSQAR2, OSQAR3} at CERN, ALPS~\cite{ALPS,ALPS2,ALPS3,ALPS4} at DESY and LIPSS~\cite{LIPSS,LIPSS2} at Jefferson Laboratories; these experiments use lasers with wavelengths $\sim \mathcal{O}(1000 \, {\rm nm})$ and powers $\lesssim \mathcal{O}(100 \, {\rm W})$. More recently, though, the use of MW-power microwave ({\it e.g.}, gyrotron and kylotron) sources has been proposed by the STAX collaboration~\cite{STAX, STAX2, STAX3, ppt_STAX, STAX4}. Independently of the hypothetical particle being searched or the particular LSW implementation, the greatest experimental challenge is the extremely low number of signal photons to be detected, since the conversion probability is $\lesssim 10^{-20}$~\cite{Ringwald}. In order to maximize the signal, the light source has to be very intense and, most importantly, the photon detector behind the wall must be as sensitive as possible.

Given the smallness of the photon mass and the expected suppression of the transmission of de Broglie-Proca waves through an optical barrier, the techniques employed in LSW experiments seem to be adequate (at least in principle) to search for massive photons in the laboratory. We analyse the transmission of massive waves normally incident upon a conductor in two cases: a semi-infinite medium (one interface) and a slab (two interfaces), the latter being more interesting in the context of LSW setups. As we shall see, it is the elusive longitudinal mode that may give us the best chance of setting upper limits on the photon rest mass, though reaching the level of other limits based on terrestrial phenomena may be challenging.

This paper is organized as follows: in Sec.~\ref{sec_matter} we review the de Broglie-Proca theory in matter. Next, we study the cases of oblique and normal incidence onto a semi-infinite medium in Secs.~\ref{sec_incidence} and~\ref{sec_rates}, respectively. In Sec.~\ref{sec_slab} we analyse the passage of de Broglie-Proca waves through a conducting slab and in Sec.~\ref{sec_exp_sens} we discuss experimental sensitivities in LSW-like setups. Our concluding remarks are presented in Sec.~\ref{sec_conclusions}. We use SI units throughout.


\section{The Proca field in matter}  \label{sec_matter}
\indent

A massive Abelian spin-1 boson, $A^\mu = (\phi/c, {\bf A})$, propagating in matter is described by the de Broglie-Proca Lagrangian~\cite{dB1, dB2, dB3, Proca1, Proca2}
\begin{equation} \label{lag_proca}
\mathcal{L} = -\frac{1}{4\mu_0} F^{\mu\nu} G_{\mu\nu} + \frac{\mu_\gamma^2}{2\mu_0} A_\mu A^\mu   - J_\mu A^\mu \, ,
\end{equation}
where $\mu_\gamma = m_\gamma c/\hbar$ is the reciprocal reduced Compton wavelength, which may be expressed as
\begin{equation}\label{eq_def_mu_gamma}
\mu_\gamma = 5.1 \times 10^{-8} \, {\rm m}^{-1} \, \left( \frac{m_\gamma}{10^{-14} \, {\rm eV/c^2}} \right)  \, .
\end{equation}
Here $J^\mu = \left( c\rho, {\bf J} \right)$ is the conserved 4-current density and the speed of light in vacuum is $c = 1/\sqrt{\varepsilon_0 \mu_0}$. The anti-symmetric field-strength tensor is $F_{\mu\nu} = \partial_\mu A_\nu - \partial_\nu A_\mu$ with the electric and magnetic fields given by $F_{0i} = {\bf E}_i/c$ and $F_{ij} = -\varepsilon_{ijk}{\bf B}_k$, respectively. The field-strength tensor $G_{\mu\nu}$ is analogous to $F_{\mu\nu}$, but with the substitutions ${\bf E} \rightarrow {\bf D}$ and ${\bf B} \rightarrow {\bf H}$, where ${\bf D} = \varepsilon {\bf E}$ and ${\bf H} = {\bf B}/\mu$ are the usual constitutive relations for homogeneous and isotropic media.


The field strengths are defined in terms of the scalar potential $\phi$ and vector potential ${\bf A}$ as usual:
\begin{equation} \label{eq_def_B_E}
{\bf B} = \nabla\times {\bf A} \quad\quad {\rm and} \quad\quad {\bf E} = -\nabla\phi - \frac{\partial {\bf A}}{\partial t}  \, ,
\end{equation}
and the inhomogeneous de Broglie-Proca equations read
\begin{subequations}
\begin{eqnarray}
\nabla\cdot {\bf D} + \frac{\mu_\gamma^2}{\mu c^2} \phi &=& \rho_f  \, , \label{eq_gauss2} \\ 
\nabla\times {\bf H} - \frac{\partial {\bf D} }{\partial t} + \frac{\mu_\gamma^2}{\mu} {\bf A} &=& {\bf J}_f \, . \label{eq_ampere2}
\end{eqnarray}
\end{subequations}
Here $\rho_f$ is the density of free charges and ${\bf J}_f = \sigma {\bf E}$ is the ohmic current. The homogeneous equations, stemming from the Bianchi identities, remain unchanged:
\begin{equation} \label{eq_homo}
\nabla\times {\bf E} + \frac{\partial {\bf B} }{\partial t} = 0  \quad\quad {\rm and} \quad\quad \nabla\cdot {\bf B} = 0 \, .
\end{equation}

We are interested in the propagation of plane waves, so let us consider the vector potential
\begin{equation} \label{eq_A}
{\bf A}({\bf r}, t) = A_0 \, e^{-i\left( \omega t - {\bf k}\cdot {\bf r} \right)} \, \hat{{\bf a}} \, ,
\end{equation}
where $A_0$ is a complex amplitude and $\hat{{\bf a}}$ is a real unit vector determining the polarization direction. The free-charge density and current are related by the continuity equation, which implies the subsidiary, Lorentz-like condition on the potentials:
\begin{equation}  \label{eq_lorenz}
\phi = \left( c^2/\omega \right) {\bf k} \cdot {\bf A} \, .
\end{equation}
Note that it is the light speed in vacuum, $c$, that appears in this relation, not the light speed in the medium. This allows us to promptly identify the three polarization states of a massive vector boson~\cite{Dombey}. A transverse polarization vector ${\bf \hat{a}}$ describes a type-I polarization, from which a type-II vector ${\bf \hat{k}} \times {\bf \hat{a}}$, also transverse, may be derived. A longitudinal polarization is perpendicular to both, pointing in the direction of ${\bf \hat{k}}$. It is worth noting that only longitudinal modes have $\phi \neq 0$.

For the sake of convenience, let us derive a few important results~\cite{Kroll1, Caccavano, Teo, Dombey}. From the equations of motion and using the subsidiary condition we find that the vector potential satisfies (making $\partial/\partial t \rightarrow -i\omega$)
\begin{equation} \label{eq_A}
\left[ \nabla^2 + \left(  \frac{n^2 \omega^2}{c^2}   - \mu_\gamma^2 \right)  \right] {\bf A} = \left( 1 - n^2 \right) \nabla \left( \nabla \cdot {\bf A}  \right) \, ,
\end{equation}
where the index of refraction is defined as
\begin{equation} \label{eq_def_n}
n = \sqrt{ \frac{\varepsilon\mu}{\varepsilon_0\mu_0} \left( 1 + i\frac{\sigma}{\varepsilon \omega} \right) } \, .
\end{equation}

The complex term related to the finite conductivity leads to attenuation. For the two transverse polarizations the term on the right-hand side of eq.~\eqref{eq_A} is zero, whereas for longitudinal polarization we may use eq.~\eqref{eq_lorenz}, so that we have (setting $\nabla \rightarrow i {\bf k}$ with $k = |{\bf k}|$)
\begin{equation}  \label{eq_DRs}
\omega^2 = \frac{c^2}{n^2} \left( k^2_{\rm T} + \mu_\gamma^2 \right)  \quad {\rm and} \quad \omega^2 = c^2 \left( k^2_{\rm L} + \frac{\mu_\gamma^2}{n^2} \right)  \, .
\end{equation}
This shows that, unless in vacuum, transverse and longitudinal polarizations have different dispersion relations. A finite conductivity affects the polarization states differently: in the limit of a perfect conductor ($|n| \rightarrow \infty$), only longitudinal modes propagate and do that at the speed of light. In this case, the electric and magnetic fields are zero, but not the potentials.

The total rate of work done by the fields in a continuous charge distribution within a volume $V$ is~\cite{Jackson}
\begin{equation}
\int_V {\bf J}_f \cdot {\bf E} \, d^3 x \, .
\end{equation}
Using eq.~\eqref{eq_ampere2} and standard vector identities, we may write this as
\begin{equation}
- \int_V \left[ \nabla \cdot {\bf S}_{\rm M} + \frac{\partial u_{\rm M}}{\partial t} - \frac{\mu_\gamma^2}{\mu} {\bf A} \cdot {\bf E}   \right] \, d^3 x \, ,
\end{equation}
where the Maxwellian Poynting vector and energy density are ${\bf S}_{\rm M} = {\bf E} \times {\bf H}$ and $u_{\rm M} = \left(  {\bf B} \cdot {\bf H} + {\bf D} \cdot {\bf E} \right)/2$, respectively. Now, using eq.~\eqref{eq_def_B_E} and the subsidiary condition, eq.~\eqref{eq_lorenz}, we obtain the statement of energy conservation for massive electrodynamics
\begin{equation}
\frac{\partial u_{\rm P}}{\partial t} = -{\bf J}_f \cdot {\bf E} - \nabla \cdot {\bf S}_{\rm P} \, ,
\end{equation}
with the Proca Poynting vector and energy density expressed in terms of fields and potentials as
\begin{subequations}
\begin{eqnarray}
{\bf S}_{\rm P} & = & {\bf E} \times {\bf H} + \frac{\mu_\gamma^2}{\mu}  \phi {\bf A} \, , \label{eq_Poynting} \\
u_{\rm P} & = & \frac{1}{2} \left(  {\bf B} \cdot {\bf H} + {\bf D} \cdot {\bf E} \right)  + \frac{\mu_\gamma^2}{2\mu c^2} \left(  \phi^2 + c^2 {\bf A}^2 \right)  . \label{eq_energy}
\end{eqnarray}
\end{subequations}
More interesting to our discussion are the respective time averages, which may be conveniently written in terms of the complex amplitudes as~\cite{Jackson}
\begin{subequations}
\begin{eqnarray}
\langle {\bf S}_{\rm P} \rangle & = & \frac{1}{2} {\rm Re}\left( {\bf E} \times {\bf H}^\star \right) + \frac{\mu_\gamma^2}{2\mu}  {\rm Re}\left( \phi {\bf A}^\star \right) \, , \label{eq_Poynting_av} \\
\langle u_{\rm P} \rangle & = & \frac{1}{4} {\rm Re}\left( {\bf B} \cdot {\bf H}^\star + {\bf D} \cdot {\bf E}^\star \right) \nonumber \\
 & + & \frac{\mu_\gamma^2}{4\mu c^2} {\rm Re} \left( |\phi|^2 + c^2 |{\bf A}|^2 \right)  \, . \label{eq_energy_av}
\end{eqnarray}
\end{subequations}
Therefore, even in regions where both electric and magnetic fields are zero ({\it e.g.}, inside a perfect conductor), there is a non-zero energy density due to the potentials.



\section{Semi-infinite medium: oblique incidence} \label{sec_incidence}
\indent

As a first step, let us consider the passage of a plane wave from a medium with $\{ \sigma_1, \varepsilon_1, \mu_1 \}$ to a medium with $\{ \sigma_2, \varepsilon_2, \mu_2 \}$. We place the flat boundary at the $x-y$ plane with the normal pointing in the $z$ direction. Requiring that the phases match at the boundary, we find that the frequencies of incident, reflected and transmitted waves are equal: $\omega_{\rm i} = \omega_{\rm r} = \omega_{\rm t} = \omega$. Analogously, assuming that the wave vectors are contained in the $x-z$ plane, their magnitudes satisfy
\begin{equation} \label{eq_ks}
k_{\rm i} \sin\theta_{\rm i} = k_{\rm r} \sin\theta_{\rm r} = k_{\rm t} \sin\theta_{\rm t} \, ,
\end{equation}
showing that $\theta_{\rm i} = \theta_{\rm r}$, since $k_{\rm i}$ is equal to $k_{\rm r}$ because the waves are propagating in the same medium.

We are interested in determining the properties of waves moving through an interface and for this it is necessary to establish the boundary conditions. Denoting components parallel and perpendicular to the interface via  the subscripts $\parallel$ and $\perp$, respectively, it can be shown that $\phi$, ${\bf A}$, ${\bf B}_{\perp}$, ${\bf E}_{\parallel}$ and ${\bf H}_{\parallel}$ are continuous across the interface~\cite{Kroll1, Teo}, though these conditions are not necessarily all independent, as we shall see shortly.

For the geometry described in the beginning of this section (with the $y$ direction pointing into the page), the incident, reflected and transmitted wave vectors are written as ${\bf k}_{\rm i} = k_{\rm i} (\sin\theta_{\rm i}, 0 , \cos\theta_{\rm i})$, ${\bf k}_{\rm r} = k_{\rm i} (\sin\theta_{\rm i}, 0 , -\cos\theta_{\rm i})$ and ${\bf k}_{\rm t} = k_{\rm t} (\sin\theta_{\rm t}, 0 , \cos\theta_{\rm t})$, respectively, cf. eq.~\eqref{eq_ks}. Furthermore, let us set the direction of the polarization vector for type-I waves along the $y$ direction, {\it i.e.}, $\hat{{\bf a}} = \hat{{\bf y}}$, parallel to the interface. For type-II waves we have $\hat{{\bf a}} = \hat{{\bf k}} \times \hat{{\bf y}}$, whereas $\hat{{\bf a}} = \hat{{\bf k}}$ for longitudinal modes. We are now ready to analyse these cases separately.

\subsection{Type-I polarization} \label{sec_type_I}
\indent

The electric and magnetic fields are (cf, eq.~\eqref{eq_def_B_E})
\begin{equation} \label{eq_EB_type_I}
{\bf E} = i\omega A_0 \hat{{\bf y}}  \quad\quad {\rm and} \quad\quad  {\bf B} = iA_0 {\bf k} \times\hat{{\bf y}} \, ,
\end{equation}
that is, the electric field is polarized perpendicular to the incidence plane. The scalar potential is zero and the vector potentials have components only along the $y$ direction, so that
\begin{equation} \label{eq_type_I_1}
A_{\rm 0i} + A_{\rm 0r} = A_{\rm 0t} \, ,
\end{equation}
which is also obtained via the continuity of ${\bf B}_{\perp}$ and ${\bf E}_{\parallel}$. The continuity of ${\bf H}_{\parallel}$ gives
\begin{equation} \label{eq_type_I_2}
\frac{k_{\rm i}\cos\theta_{\rm i}}{\mu_1} \left( A_{\rm 0i} - A_{\rm 0r} \right) = \frac{k_{\rm t}\cos\theta_{\rm t}}{\mu_2} A_{\rm 0t} \, .
\end{equation}

Combining these results in light of eq.~\eqref{eq_ks} we finally obtain~\cite{Caccavano}
\begin{subequations}
\begin{eqnarray}
\frac{A_{\rm 0r}}{A_{\rm 0i}} & = & \frac{ k_{\rm i}\cos\theta_{\rm i} - \frac{\mu_1}{\mu_2}\sqrt{ k_{\rm t}^2 - k_{\rm i}^2\sin^2\theta_{\rm i}} }{ k_{\rm i}\cos\theta_{\rm i} + \frac{\mu_1}{\mu_2}\sqrt{ k_{\rm t}^2 - k_{\rm i}^2\sin^2\theta_{\rm i}} }  \, , \label{eq_type_I_r}  \\
\frac{A_{\rm 0t}}{A_{\rm 0i}} & = & \frac{ 2k_{\rm i}\cos\theta_{\rm i}  }{ k_{\rm i}\cos\theta_{\rm i} + \frac{\mu_1}{\mu_2}\sqrt{ k_{\rm t}^2 - k_{\rm i}^2\cos\theta_{\rm i}} } \label{eq_type_I_t} \, ,
\end{eqnarray} 
\end{subequations}
with the wave numbers determined by the transverse dispersion relation for each medium, cf. eq.~\eqref{eq_DRs}. These equations are the generalization of the Fresnel equations for massive electrodynamics -- also here there is no Brewster angle for this polarization under normal circumstances ($\mu_1 \approx \mu_2$)~\cite{Jackson}. Equations~\eqref{eq_type_I_r} and~\eqref{eq_type_I_t} reduce to the results of Maxwell's theory in the massless limit.

\subsection{Type-II polarization} \label{sec_type_II}
\indent

The electric and magnetic fields are 
\begin{equation} \label{eq_EB_type_II}
{\bf E} = i\omega A_0 \hat{{\bf k}} \times \hat{{\bf y}}  \quad\quad {\rm and} \quad\quad  {\bf B} = -iA_0 k\hat{{\bf y}} 
\end{equation}
with the electric field polarized parallel to the incidence plane. Here $\phi = 0$ and the continuity of the vector potential (only non-zero in the $x$ and $z$ directions) gives
\begin{subequations}
\begin{eqnarray}
\left( A_{\rm 0i} + A_{\rm 0r} \right) \sin\theta_{\rm i} & = &  A_{\rm 0t} \sin\theta_{\rm t} \, , \label{eq_type_II_1}  \\
\left( A_{\rm 0i} - A_{\rm 0r} \right) \cos\theta_{\rm i} & = &  A_{\rm 0t} \cos\theta_{\rm t} \label{eq_type_II_2} \, ,
\end{eqnarray}
\end{subequations}
with the continuity of ${\bf E}_\parallel$ being redundant to eq.~\eqref{eq_type_II_2}. Now, the continuity of ${\bf H}_\parallel$ implies
\begin{equation} \label{eq_type_II_3}
A_{\rm 0i} + A_{\rm 0r} = \frac{\mu_1 k_{\rm t}}{\mu_2 k_{\rm i}}  A_{\rm 0t} \, ,
\end{equation}
which is inconsistent with eq.~\eqref{eq_type_II_1} in light of eq.~\eqref{eq_ks}. This indicates that pure type-II waves, unlike type-I waves, are incompatible with the boundary conditions.

\subsection{Longitudinal polarization} \label{sec_type_long}
\indent

For this polarization mode we have ${\bf B} = {\bf H} = 0$. The scalar potential is given by (cf. eq.~\eqref{eq_lorenz})
\begin{equation} \label{eq_phi_long}
\phi = \frac{c^2 k}{\omega} A_0 \, ,
\end{equation}
whose continuity implies that
\begin{equation} \label{eq_A_long_1}
A_{\rm 0i} + A_{\rm 0r} = \frac{k_{\rm t}}{k_{\rm i}}  A_{\rm 0t}  \, .
\end{equation}
Using the longitudinal dispersion relation, cf. eq.~\eqref{eq_DRs}, the electric field in each medium is
\begin{equation} \label{eq_E_long}
{\bf E} = i\omega \frac{\eta^2 }{n^2} A_0 \hat{{\bf k}} \, ,
\end{equation}
where the dimensionless parameter $\eta$ is given by
\begin{equation} \label{eq_def_eta}
\eta = \frac{\mu_\gamma c}{\omega} = \frac{m_\gamma c^2}{\hbar\omega} \, .
\end{equation}
The continuity of its components parallel to the flat interface leads to
\begin{equation} \label{eq_A_long_2}
A_{\rm 0i} + A_{\rm 0r} = \frac{ n_1^2 \sin\theta_{\rm t} }{ n_2^2 \sin\theta_{\rm i} } A_{\rm 0t}  \, ,
\end{equation}
which is at odds with eq.~\eqref{eq_A_long_1}. Therefore, purely longitudinal waves, like type-II waves, are incompatible with the boundary conditions.

\subsection{Mixed-type polarization} \label{sec_type_II_long}
\indent

As shown above and reported in Refs.~\cite{Dombey, Caccavano, Teo}, type-II and longitudinal polarizations do not fulfil the boundary conditions separately. We therefore consider a linear superposition of these two modes (identified by the superscripts $^{\rm (II)}$ and $^{\rm (L)}$), so that the potentials become\footnote{A second, orthogonal linear combination may be constructed, but we shall not work out its details explicitly.}
\begin{subequations}
\begin{eqnarray}
\phi & = & \phi_{\rm II} + \phi_{\rm L} = \frac{c^2 k_{\rm L}}{\omega} A_0^{\rm (L)} \label{eq_mix_1} \, , \\
{\bf A} & = & {\bf A}_{\rm II} + {\bf A}_{\rm L} = A_0^{\rm (II)} \hat{{\bf k}} \times \hat{{\bf y}} + A_0^{\rm (L)} \hat{{\bf k}} \, . \label{eq_mix_2}
\end{eqnarray}
\end{subequations}
The different modes have different dispersion relations and each field (or potential) is multiplied by an exponential factor $\exp(i {\bf k}_{\rm L} \cdot {\bf r})$ or $\exp(i {\bf k}_{\rm T} \cdot {\bf r})$. These are not identical in general, but the matching conditions of the fields (and potentials) along the flat interface also induces a matching of these exponential factors. This occurs at $z = 0$ and must be valid for all the $x-y$ plane, in particular at $x = y = 0$. Therefore, for a single flat interface, the different exponential factors related to the wave numbers may be safely ignored. This is not so for two (or more) interfaces, cf. Sec.~\ref{sec_slab}.

The continuity of $\phi$ and ${\bf A}$ lead to 
\begin{subequations}
\begin{eqnarray}
A_{\rm 0i}^{\rm (L)} + A_{\rm 0r}^{\rm (L)} &=& \frac{k_{\rm t}^{\rm L}}{k_{\rm i}^{\rm L}}  A_{\rm 0t}^{\rm (L)}  \, , \label{eq_mix_3}  \\
\left( A_{\rm 0i}^{\rm (II)} + A_{\rm 0r}^{\rm (II)} \right) \sin\theta_{\rm i} &+& \left( A_{\rm 0i}^{\rm (L)} - A_{\rm 0r}^{\rm (L)} \right) \cos\theta_{\rm i} = \nonumber \\
= A_{\rm 0t}^{\rm (II)} \sin\theta_{\rm t} &+& A_{\rm 0t}^{\rm (L)} \cos\theta_{\rm t}  \, ,\label{eq_mix_4} \\
\left( A_{\rm 0i}^{\rm (II)} - A_{\rm 0r}^{\rm (II)} \right) \cos\theta_{\rm i} &-& \left( A_{\rm 0i}^{\rm (L)} + A_{\rm 0r}^{\rm (L)} \right) \sin\theta_{\rm i} = \nonumber \\
= A_{\rm 0t}^{\rm (II)} \cos\theta_{\rm t} &-& A_{\rm 0t}^{\rm (L)} \sin\theta_{\rm t}  \, , \label{eq_mix_5}
\end{eqnarray}
\end{subequations}
whereas the continuity of ${\bf E}_\parallel$ gives (cf. eq.~\eqref{eq_def_eta})
\begin{eqnarray}
\left( A_{\rm 0i}^{\rm (II)} - A_{\rm 0r}^{\rm (II)} \right) \cos\theta_{\rm i} &-& \left( A_{\rm 0i}^{\rm (L)} + A_{\rm 0r}^{\rm (L)} \right) (\eta/n_1)^2 \sin\theta_{\rm i} = \nonumber \\
= A_{\rm 0t}^{\rm (II)} \cos\theta_{\rm t} &-& A_{\rm 0t}^{\rm (L)} (\eta/n_2)^2 \sin\theta_{\rm t} \, .  \label{eq_mix_6}
\end{eqnarray}
Finally, from ${\bf B}_\perp$ we do not learn anything, since it is identically zero for the longitudinal mode and ${\bf B}$ points entirely in the $y$ direction for the type-II polarization. Now, ${\bf H}_\parallel$ is equally zero for the longitudinal mode, but for the type-II polarization we find
\begin{eqnarray}
A_{\rm 0i}^{\rm (II)} + A_{\rm 0r}^{\rm (II)} &=& \frac{\mu_1 k_{\rm t}^{\rm T} }{\mu_2 k_{\rm i}^{\rm T} }  A_{\rm 0t}^{\rm (II)}   \, . \label{eq_mix_7}
\end{eqnarray}

Let us now determine the ratios of reflected and transmitted amplitudes relative to the incident ones. Subtracting eq.~\eqref{eq_mix_6} from eq.~\eqref{eq_mix_5} eliminates the type-II terms and, using eq.~\eqref{eq_ks}, we recover eq.~\eqref{eq_mix_3}. We are then allowed to ignore eq.~\eqref{eq_mix_6} in favor of eq.~\eqref{eq_mix_3}. The solution to the system of equations is
\begin{widetext}
\begin{subequations}
\begin{eqnarray}
\frac{A_{\rm 0r}^{\rm (L)}}{A_{\rm 0i}^{\rm (L)} } & = & \frac{ (\alpha + \beta) \sin\theta_i \sin \theta_t + \alpha\beta \cos 2\theta_i +\cos \theta_i \left[ (\alpha - \beta) \cos \theta_t -2 \alpha \rho^{-1} 
\sin\theta_t \right] + \alpha\beta \rho^{-1} \sin 2\theta_i -1}{(\alpha + \beta) \cos (\theta_i +\theta_t)+\alpha \beta+1} \label{eq_mix_r_L} \; ,
\\
\frac{A_{\rm 0r}^{\rm (II)}}{A_{\rm 0i}^{\rm (II)} } &=& \frac{ (\beta - \alpha) \cos\theta_i \cos\theta_t + \alpha\beta \cos 2\theta_i + \sin\theta_t \left[ (\alpha + \beta) \sin\theta_i +2\beta \rho \cos\theta_i \right] - \alpha\beta \rho \sin 2\theta_i  - 1}{(\alpha + \beta) \cos (\theta_i +\theta_t)+\alpha \beta+1} \label{eq_mix_r_II} \; ,
\\
\frac{A_{\rm 0t}^{\rm (L)}}{A_{\rm 0i}^{\rm (L)} } &=& \frac{2 \cos\theta_i \left[ \rho^{-1} (\beta\sin\theta_i -\sin\theta_t ) + \beta\cos\theta_i +\cos\theta_t  \right] }{(\alpha + \beta) \cos (\theta_i +\theta_t)+\alpha \beta+1} \label{eq_mix_t_L} \; ,
\\
\frac{A_{\rm 0t}^{\rm (II)}}{A_{\rm 0i}^{\rm (II)} } &=& \frac{2 \cos\theta_i \left[ \rho (\sin\theta_t -\alpha \sin\theta_i)+\alpha \cos\theta_i +\cos\theta_t \right] }{(\alpha + \beta) \cos (\theta_i +\theta_t)+\alpha \beta+1} \label{eq_mix_t_II} \; ,
\end{eqnarray}
\end{subequations}
\end{widetext}
where we defined the following parameters
\begin{equation} \label{eq_def_alpha_beta}
\alpha = \frac{ k_{\rm t}^{\rm L} }{ k_{\rm i}^{\rm L} } \quad {\rm and} \quad \beta = \frac{ \mu_1 k_{\rm t}^{\rm T} }{ \mu_2 k_{\rm i}^{\rm T} }  \, .
\end{equation}
This shows that a linear combination of type-II and longitudinal modes is compatible with the boundary conditions at the interface. Furthermore, we defined the ratio of the longitudinal and transverse type-II amplitudes as
\begin{equation} \label{eq_def_rho}
\rho = \frac{ A_{\rm 0i}^{\rm (L)} }{ A_{\rm 0i}^{\rm (II)} } \, .
\end{equation}
Here $\rho$ is a free parameter. Experimentally, however, there is so far no clear sign of longitudinal photons and we may safely assume that $|\rho| \ll 1$, cf. Sec.~\ref{sec_exp_sens}.

\section{Semi-infinite medium: reflection and transmission rates at normal incidence} \label{sec_rates}
\indent

The intensity of a monochromatic wave is given by the average power per unit area. This may be expressed as $I = \langle |{\bf S}_{\rm P} \cdot \hat{{\bf z}} | \rangle$, cf. eq.~\eqref{eq_Poynting_av}, and the reflection and transmission rates are
\begin{equation} \label{eq_R_T}
R = \frac{ I_{\rm r} }{ I_{\rm i} } \quad\quad {\rm and} \quad\quad T = \frac{ I_{\rm t} }{ I_{\rm i} }  \, .
\end{equation}
For normal incidence ($\theta_{\rm i} = \theta_{\rm t} = 0$) we have
\begin{equation} \label{eq_I}
I  = \frac{1}{2\mu} {\rm Re}\left[ \left( {\bf E} \times {\bf B}^\star \right) \cdot \hat{{\bf z}} \right] + \frac{\mu_\gamma^2}{2\mu}  {\rm Re} \left[ \phi \left( {\bf A}^\star \cdot \hat{{\bf z}} \right) \right] \, ,
\end{equation}
and we may now study the transmission and reflection rates for type-I and mixed-type waves.



\subsection{Type-I polarization} \label{sec_type_I_rate}
\indent


For this polarization we have $\phi = 0$, so only the first term in eq.~\eqref{eq_I} contributes. From eq.~\eqref{eq_EB_type_I} we find
\begin{equation} \label{eq_I_1}
I_{\rm j} = \frac{\omega}{2\mu} |A_{\rm 0j}|^2 {\rm Re} (k_{\rm j})  \, ,
\end{equation}
where $j = i,r,t$ and we are omitting the subscript $_{\rm T}$ for the transverse wave number, cf. eq.~\eqref{eq_DRs}. From eqs.~\eqref{eq_type_I_r} and~\eqref{eq_type_I_t} we get
\begin{subequations}
\begin{eqnarray}
R & = & \bigg| \frac{A_{\rm 0r}}{A_{\rm 0i}} \bigg|^2 = \frac{|\mu_2 k_{\rm i} - \mu_1 k_{\rm t}|^2}{|\mu_2 k_{\rm i} + \mu_1 k_{\rm t}|^2}  \, , \label{eq_I_2} \\
T & = & \bigg| \frac{A_{\rm 0t}}{A_{\rm 0i}} \bigg|^2 \frac{{\rm Re}(k_{\rm t})}{{\rm Re}(k_{\rm i})} =  \frac{4\mu_2^2 |k_{ \rm i}|^2}{|\mu_2 k_{\rm i} + \mu_1 k_{\rm t}|^2} \frac{{\rm Re}(k_{\rm t})}{{\rm Re}(k_{\rm i})}  \, . \label{eq_I_3}
\end{eqnarray}
\end{subequations}
These results are general, but now we specialize them to waves moving from vacuum $\{ \sigma_1 = 0, \varepsilon_1 = \varepsilon_0, \mu_1 = \mu_0 \}$, into a nonpermeable medium $\{ \sigma_2 = \sigma, \varepsilon_2 = \varepsilon, \mu_2 = \mu_0 \}$.

Under the conditions stated above, we find that
\begin{equation} \label{eq_amp_r}
\frac{A_{\rm 0r}}{A_{\rm 0i}} = \frac{1 -  \beta }{1 +  \beta}  \, ,
\end{equation}
with $\beta$ given by eq.~\eqref{eq_def_alpha_beta} with $\mu_1 = \mu_2 = \mu_0$. From eq.~\eqref{eq_DRs} we may expand $\beta$ as
\begin{equation}\label{eq_ratio}
\beta = n_{\rm t} \left( 1 + \frac{1}{2}\eta^2 \right) + \mathcal{O} \left(\eta^4 \right) \, ,
\end{equation}
where the photon mass is implicitly contained in $\eta$ as defined by eq.~\eqref{eq_def_eta} and a good conductor is assumed. Here we neglect terms of order $\mathcal{O} \left(\eta^2/n_{\rm t}^2 \right)$, since the effects of the conductivity are already captured by the pre-factor.

From now on we work in the limit of high conductivity, where $\sigma \gg \varepsilon_0\omega$. In this case, the index of refraction becomes (cf. eq.~\eqref{eq_def_n})
\begin{equation} \label{eq_n} 
n_{\rm t}  \approx  (1 + i)\frac{c}{\omega\delta_0} \, ,
\end{equation}
where $\delta_0 = \sqrt{ 2 / \mu_0 \sigma \omega}$ is the skin depth for a good conductor in the massless limit~\cite{Jackson}. Taking copper with $\sigma = 5.9 \times 10^7 \, {\rm S/m}$ as a reference, it may be conveniently expressed as
\begin{equation}  \label{eq_def_skin_numeric}
\delta_0 =  3.8 \, {\rm nm}  \left( \frac{5.9 \times 10^7 \, {\rm S/m} }{\sigma} \right)^{1/2} \left( \frac{\lambda_{\rm source}}{1000 \, {\rm nm} } \right)^{1/2}  \, .
\end{equation}
Moving on, in light of eq.~\eqref{eq_n} we may write $\beta \approx (1 + i) \overline{\beta}$ with 
\begin{equation}  \label{eq_def_beta_bar}
\overline{\beta} = \frac{c}{\omega\delta_0} \left( 1 + \frac{1}{2}\eta^2 \right)  \, ,
\end{equation}
where we once again neglected terms of order $\mathcal{O}(\eta^2/n_{\rm t}^2)$. Inserting these results into eq.~\eqref{eq_amp_r} we are able to write the reflection rate as
\begin{equation} \label{eq_R_rate_I}
R \approx 1 - \frac{2\omega\delta_0}{c}  \left( 1 - \frac{1}{2}\eta^2  \right) + 2\left( \frac{\omega\delta_0}{c} \right)^2 \left( 1 - \eta^2  \right)  \, .
\end{equation}
It is clear from this result that the presence of a finite photon mass slightly enhances the already high reflection rate of a good conductor, that is, $R(\eta) > R(0)$.

Let us now move on to the transmission rate, eq.~\eqref{eq_I_3}, which we write as $T = T_1 T_2$ with ($\mu_1 = \mu_2 = \mu_0$)
\begin{equation} \label{eq_T12}
T_1 = \frac{4}{|1 + \beta|^2} \quad {\rm and} \quad T_2 = \frac{{\rm Re}(k_{\rm t})}{{\rm Re}(k_{\rm i})} \, .
\end{equation}
Returning to the dispersion relation for a type-I wave, cf. eq.~\eqref{eq_DRs}, we see that $k_{\rm i}$ is real, {\it i.e.}, ${\rm Re}(k_{\rm i}) = k_{\rm i}$, since $n_{\rm i} = 1$ for medium 1, vacuum. For the transmitted wave the wave number is complex due to the finite conductivity term. We thus have ${\rm Re}(k_{\rm t})/{\rm Re}(k_{\rm i}) = {\rm Re} \left( k_{\rm t}/k_{\rm i} \right)$, that is (cf. eqs.~\eqref{eq_ratio} and~\eqref{eq_n})
\begin{equation} 
\frac{{\rm Re}(k_{\rm t})}{{\rm Re}(k_{\rm i})} = {\rm Re} \left( \beta \right) \, ,
\end{equation}
implying that $T_2 = \overline{\beta}$. Again taking the limit of a good conductor, we finally obtain
\begin{equation} \label{eq_T_rate_I}
T \approx \frac{2\omega\delta_0}{c}  \left( 1 - \frac{1}{2}\eta^2  \right) - 2\left( \frac{\omega\delta_0}{c} \right)^2 \left( 1 - \eta^2  \right)  \, .
\end{equation}
A finite photon mass causes the transmission to slightly decrease, that is, $T(\eta) < T(0)$. This reduction is the same as the increment in the reflection rate, so $R + T \approx 1$, as expected.

\subsection{Mixed-type polarizations} \label{sec_type_II_long_rate}
\indent

Let us now consider mixed-type waves, where longitudinal and type-II waves are combined in a linear superposition. Once more, here we focus on waves moving from vacuum $\{ \sigma_1 = 0, \varepsilon_1 = \varepsilon_0, \mu_1 = \mu_0 \}$ into a highly conducting, nonpermeable medium $\{ \sigma_2 = \sigma, \varepsilon_2 = \varepsilon, \mu_2 = \mu_0 \}$. From eq.~\eqref{eq_I} we have
\begin{equation}\label{eq_I_mixed} 
I_{\rm j} = \frac{\omega}{2\mu_0} | A_{\rm 0j}^{\rm (II)} |^2 \, {\rm Re}(k_{\rm j}^{\rm T})  +  \frac{\omega \eta^2}{2\mu_0} | A_{\rm 0j}^{\rm (L)} |^2 \,  {\rm Re}(k_{\rm j}^{\rm L})  \, ,
\end{equation}
with $j = i,r,t$ and the wave number of type-II waves being denoted by $k_j^{\rm T}$. The transverse type-II component only contributes to the $\sim {\bf E} \times {\bf B}$ term, since $\phi = 0$, whereas the longitudinal mode only appears in the $\sim \phi {\bf A}$ term, which is strongly suppressed by a factor of $\eta^2$. The reflection and transmission rates become
\begin{subequations}
\begin{eqnarray}
R & = & \frac{ | A_{\rm 0r}^{\rm (II)} |^2 \, {\rm Re}(k_{\rm i}^{\rm T}) + \eta^2 | A_{\rm 0r}^{\rm (L)} |^2 \, {\rm Re}(k_{\rm i}^{\rm L})  }{ | A_{\rm 0i}^{\rm (II)} |^2 \, {\rm Re}(k_{\rm i}^{\rm T}) + \eta^2 | A_{\rm 0i}^{\rm (L)} |^2 \, {\rm Re}(k_{\rm i}^{\rm L}) } \, , \label{eq_R_mixed} \\
T & = & \frac{ | A_{\rm 0t}^{\rm (II)} |^2 \, {\rm Re}(k_{\rm t}^{\rm T}) + \eta^2 | A_{\rm 0t}^{\rm (L)} |^2 \, {\rm Re}(k_{\rm t}^{\rm L})  }{ | A_{\rm 0i}^{\rm (II)} |^2 \, {\rm Re}(k_{\rm i}^{\rm T}) + \eta^2 | A_{\rm 0i}^{\rm (L)} |^2 \, {\rm Re}(k_{\rm i}^{\rm L}) } \, , \label{eq_T_mixed} 
\end{eqnarray}
\end{subequations}
where we used that $k_{\rm r} = k_{\rm i}$ for both transverse and longitudinal modes propagating in vacuum.

To progress further, we take the limit of normal incidence in eqs.~\eqref{eq_mix_r_L}-\eqref{eq_mix_t_II}, which neatly reduce to
\begin{subequations}
\begin{eqnarray}
\frac{A_{\rm 0r}^{\rm (L)}}{A_{\rm 0i}^{\rm (L)} } & = & \frac{\alpha - 1}{\alpha + 1}  \quad {\rm and} \quad \frac{A_{\rm 0t}^{\rm (L)}}{A_{\rm 0i}^{\rm (L)} } = \frac{2}{\alpha + 1}  \label{eq_mix_rt_L_2} \; ,
\\
\frac{A_{\rm 0r}^{\rm (II)}}{A_{\rm 0i}^{\rm (II)} } &=& \frac{\beta - 1}{\beta + 1} \quad {\rm and} \quad \frac{A_{\rm 0t}^{\rm (II)}}{A_{\rm 0i}^{\rm (II)} } = \frac{2}{\beta + 1} \label{eq_mix_rt_II_2}  \; ,
\end{eqnarray}
\end{subequations}
with $\alpha = k_{\rm t}^{\rm L} / k_{\rm i}^{\rm L}$ and $\beta = k_{\rm t}^{\rm T} / k_{\rm i}^{\rm T}$ established in eq.~\eqref{eq_def_alpha_beta}. We have already determined the parameter $\beta$, cf. eqs.~\eqref{eq_ratio} and~\eqref{eq_def_beta_bar}, but we may follow an analogous strategy to evaluate $\alpha$. Again working in the limit of good conductivity, we find 
\begin{equation} \label{eq_ratio_L}
\alpha = 1 + \frac{1}{2}\eta^2 \left[ 1 + \frac{i}{2} \left( \frac{\omega\delta_0}{c} \right)^2 \right] + \mathcal{O} \left( \eta^4 \right) \, ,
\end{equation}
where we used that $\eta^2 (\omega\delta_0/c)^2 \gg \eta^4$ for the range of photon masses and frequencies of interest. In fact, we have
\begin{equation}  \label{eq_def_gamma}
\frac{\omega\delta_0}{c} = 2.4 \times 10^{-2} \left( \frac{5.9 \times 10^7 \, {\rm S/m} }{\sigma} \right)^{1/2} \left( \frac{1000 \, {\rm nm} }{\lambda_{\rm source}} \right)^{1/2} \, 
\end{equation}
and, from eq.~\eqref{eq_def_eta}, we find that
\begin{eqnarray}
\eta & = & 8.1 \times 10^{-15} \, \left( \frac{m_\gamma}{10^{-14} \, {\rm eV/c^2}} \right) \left( \frac{ \lambda_{\rm source} }{ 1000 \, {\rm nm} } \right) \, .  \label{eq_def_eta_numeric}
\end{eqnarray}
The approximation is thus valid for $\lambda_{\rm source} \lesssim 100$~m.

The imaginary term in eq.~\eqref{eq_ratio_L} is small, but, contrary to the approximation used to determine $\beta$, cf. eq.~\eqref{eq_ratio}, here we choose to keep it, since it is the leading-order contribution from the conductivity. Its smallness attests to the fact that longitudinal waves barely experience any attenuation when traversing a conductor. Keeping only leading-order terms in $\eta$, we find
\begin{eqnarray}
\Bigg| \frac{A_{\rm 0r}^{\rm (L)}}{A_{\rm 0i}^{\rm (L)} } \Bigg|^2 & \approx & \frac{1}{16} \eta^4  \quad {\rm and} \quad \Bigg| \frac{A_{\rm 0t}^{\rm (L)}}{A_{\rm 0i}^{\rm (L)} } \Bigg|^2 \approx 1 - \frac{1}{2}\eta^2   \label{eq_mix_rt_L_3} \; .
\end{eqnarray}
These expressions further display the almost complete lack of attenuation of the longitudinal modes.

For incident and reflected waves we have $k_{\rm i}^{\rm L} = k_{\rm i}^{\rm T} = k_{\rm i}$, since $n_{\rm i} = 1$. We may write $R = R_1 / R_2$ with $R_1$ being eq.~\eqref{eq_R_rate_I} plus a term $\sim |\rho|^2 \eta^6$, which we promptly discard. The denominator is $R_2 =  1 + |\rho|^2 \eta^2$, with $\rho$ defined by eq.~\eqref{eq_def_rho}. Ignoring terms of order $\mathcal{O}(\eta^4)$ and higher, we obtain
\begin{eqnarray}
R & \approx & 1 - \frac{2\omega\delta_0}{c}  \left( 1 - \frac{1}{2}\eta^2  \right) + 2\left( \frac{\omega\delta_0}{c} \right)^2 \left( 1 - \eta^2  \right)  \nonumber \\
& - &   |\rho|^2 \eta^2 \left[ 1 - \frac{2\omega\delta_0}{c} + 2\left( \frac{\omega\delta_0}{c} \right)^2 \right]  \, . \label{eq_R_mixed_final}
\end{eqnarray} 
The longitudinal component reduces the reflectivity, whereas the transverse type-II component increases it. Nonetheless, the overall effect of a finite photon mass is not obvious without knowledge of $|\rho|$. Neglecting second-order terms in $\gamma = \omega\delta_0/c$, eq.~\eqref{eq_R_mixed_final} may be written as
\begin{equation}
R(\eta) \approx R(0) + \eta^2 \left[ 2\gamma \left( |\rho|^2 + \frac{1}{2} \right) - |\rho|^2 \right] \, .
\end{equation}
This shows that $R(\eta) > R(0)$, as in the case of type-I waves, only if $|\rho|^2 \lesssim \gamma/(1 - 2\gamma)$.

For the transmission rate we first evaluate the ratios of the real parts of the wave numbers:
\begin{subequations}
\begin{eqnarray}
\frac{ {\rm Re}(k_{\rm t}^{\rm T}) }{ {\rm Re}(k_{\rm i}) } & \approx & {\rm Re}(n_{\rm t}) \left(  1 + \frac{1}{2}\eta^2 \right) \, , \\
\frac{ {\rm Re}(k_{\rm t}^{\rm L}) }{ {\rm Re}(k_{\rm i}) } & \approx & 1 + \frac{1}{2}\eta^2 \, .
\end{eqnarray}
\end{subequations}
The last result stems from eq.~\eqref{eq_ratio_L}. Again assuming good conductivity, we have ${\rm Re}(n_{\rm t}) \approx c/\omega\delta_0$, so that $T = T_1/T_2$ with $T_1$ identical to eq.~\eqref{eq_T_rate_I} plus $|\rho|^2 \eta^2$, whereas $T_2 = 1 + |\rho|^2 \eta^2$. Ignoring terms of order $\mathcal{O}(\eta^4)$ and higher, we finally get
\begin{eqnarray}
T & \approx & \frac{2\omega\delta_0}{c}  \left( 1 - \frac{1}{2}\eta^2  \right) - 2\left( \frac{\omega\delta_0}{c} \right)^2 \left( 1 - \eta^2  \right)  \nonumber \\
& + &   |\rho|^2 \eta^2 \left[ 1 - \frac{2\omega\delta_0}{c} + 2\left( \frac{\omega\delta_0}{c} \right)^2 \right]  \, , \label{eq_T_mixed_final}
\end{eqnarray}
satisfying $R + T \approx 1$, as expected. Again, $T(\eta) < T(0)$, as for type-I waves, only if $|\rho|^2 \lesssim \gamma/(1 - 2\gamma)$ with $\gamma = \omega\delta_0/c$.

\section{Passage through a conducting slab at normal incidence} \label{sec_slab}
\indent

Here we consider a slab of thickness $\mathcal{D}$ made of a nonpermeable material characterized by $\{ \sigma_2 = \sigma \gg \varepsilon_0\omega, \varepsilon_2 = \varepsilon, \mu_2 = \mu_0 \}$ surrounded by vacuum. There are now two interfaces at which waves will be reflected and transmitted: the first is located at $z = 0$ and the second at $z = \mathcal{D}$.

On the left side ($z < 0$) incident and reflected waves have amplitudes $A_{\rm 0i}$ and $A_{\rm 0r}$, respectively. Within the slab ($0 \leq z \leq \mathcal{D}$)  transmitted waves propagate to the right ($A_{\rm 0+}$) and the reflected (at the second interface) to the left ($A_{\rm 0-}$). On the right side ($z > \mathcal{D}$), there are only waves propagating to the right with amplitude $A_{\rm 0t}$. As usual, all amplitudes carry a $\exp(i {\bf k} \cdot {\bf r})$ dependence; waves moving in the positive $z$ direction have ${\bf k} = + k \hat{{\bf z}}$, whereas those moving in the opposite direction have ${\bf k} = - k \hat{{\bf z}}$.



\subsection{Type-I polarization} \label{sec_type_I_slab}
\indent

At the first interface ($z = 0$), from the continuity of the vector potential, we have
\begin{equation} 
A_{\rm 0i} + A_{\rm 0r} = A_{\rm 0+} + A_{\rm 0-} \, ,
\end{equation}
which is reproduced by the continuity of ${\bf E}_\parallel$, whereas ${\bf B}_\perp = 0$ and $\phi = 0$ are trivial. From ${\bf H}_\parallel$ continuous we obtain
\begin{equation} 
A_{\rm 0i} - A_{\rm 0r} = \frac{k'}{k} \left( A_{\rm 0+} - A_{\rm 0-} \right) \, ,
\end{equation}
where we set $k_{\rm i} = k_{\rm r} = k_{\rm t} = k$ and $k_{\rm +} = k_{\rm -} = k'$. Applying the boundary conditions to the second interface and letting $\xi = k' \mathcal{D}$, we have
\begin{subequations}
\begin{eqnarray}
A_{\rm 0+}e^{i\xi} + A_{\rm 0-}e^{-i\xi} & = & A_{\rm 0t} \, , \\
A_{\rm 0+}e^{i\xi} - A_{\rm 0-}e^{-i\xi} & = & \frac{k}{k'} A_{\rm 0t} \, .
\end{eqnarray}
\end{subequations}

We are not interested in the amplitudes within the slab, $A_{\rm 0\pm}$, so we eliminate them in favor of the rest. There are four equations for four variables, since $A_{\rm 0i}$ is supposedly known. Identifying $\beta = k'/k$, we find
\begin{subequations}
\begin{eqnarray}
\frac{A_{\rm 0r}}{A_{\rm 0i}} & = & - \frac{ (1 - \beta^2)(1 - e^{2i\xi} ) }{e^{2i\xi} (1 - \beta)^2 - (1 + \beta)^2}  \, , \label{eq_ratio_r_I_slab} \\
\frac{A_{\rm 0t}}{A_{\rm 0i}} & = & - \frac{4\beta e^{i\xi}}{e^{2i\xi} (1 - \beta)^2 - (1 + \beta)^2}  \label{eq_ratio_t_I_slab}  \, .
\end{eqnarray}
\end{subequations}
Under the assumption of good conductivity, we may write $\xi \approx n\mathcal{D}\omega/c$, but using eq.~\eqref{eq_n}, we have $\xi \approx (1 + i)\overline{\xi}$ with
\begin{equation} \label{eq_def_xi}
\overline{\xi} = \frac{\mathcal{D}}{\delta_0} + \mathcal{O}(\eta^2/n^2) \, .
\end{equation}

The real part of $\xi$ will give rise to an oscillating term, whereas its imaginary part will result in an exponential suppression factor $\exp(-\mathcal{D}/\delta_0)$, showing that, for the transverse modes, the associated skin depth is essentially the same as in massless electrodynamics~\cite{Jackson}. The reflection and transmission rates are given by
\begin{equation} \label{eq_rates_slab_type_I}
R = \frac{|z_1|^2}{ |z_2|^2} \quad {\rm and} \quad T = \frac{|z_3|^2}{ |z_2|^2}  \, ,
\end{equation}
where
\begin{subequations}
\begin{eqnarray}
| z_1 |^2 & = &  \left( 1 + 4\overline{\beta}^4  \right) \left( 1 - 2\lambda\cos(2\overline{\xi}) + \lambda^2  \right) \, , \label{eq_z1} \\
| z_2 |^2 & = &  \left[ 1 + 2\overline{\beta} \left( 1 + \overline{\beta} \right)  \right]^2 + \lambda^2 \left[ 1 - 2\overline{\beta} \left(1 - \overline{\beta} \right)  \right]^2   \nonumber \\ 
& - & 2\lambda \left( 1 - 8\overline{\beta}^2 + 4\overline{\beta}^4 \right) \cos(2\overline{\xi})  \nonumber \\ 
& - & 8\lambda \overline{\beta} \left( 1 - 2\overline{\beta}^2 \right) \sin(2\overline{\xi})  \, , \label{eq_z2}  \\
| z_3 |^2 & = &  32 \lambda \overline{\beta}^2 \, , \label{eq_z3} 
\end{eqnarray}
\end{subequations}
with $\overline{\beta}$ defined in eq.~\eqref{eq_def_beta_bar} and we used that ${\rm Re}(k_{\rm t}) / {\rm Re}(k_{\rm i}) = 1$, since incident and transmitted waves are propagating in the same medium. Here we defined
\begin{equation}  \label{eq_def_lambda}
\lambda = \exp(-2\overline{\xi})  \, ,
\end{equation}
which is in general very small: for $\mathcal{D} = 2\delta_0$ we have $\lambda = 0.02$, going down to $\lambda = 2.1 \times 10^{-9}$ for $\mathcal{D} = 10\delta_0$. This steep decline with increasing thickness will effectively suppress the oscillatory character of the rates.

\begin{figure}[t!]
\begin{minipage}[b]{1.0\linewidth}
\includegraphics[width=\textwidth]{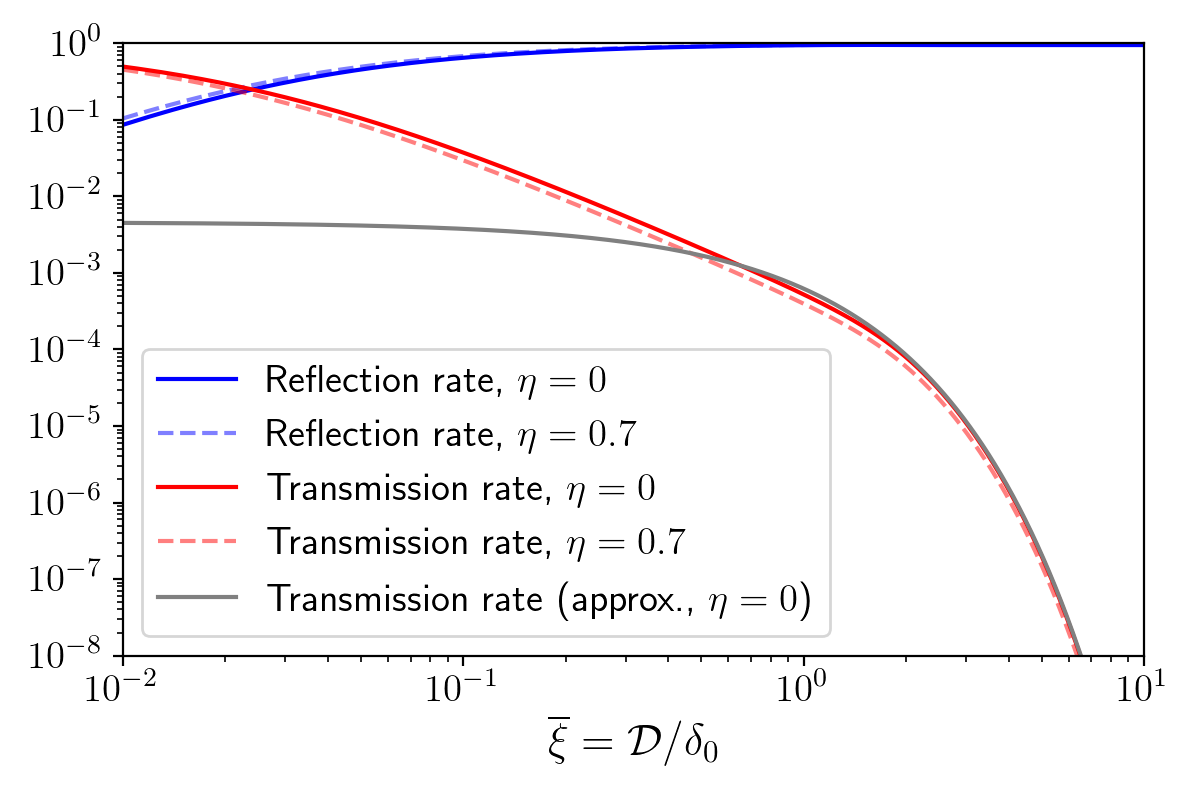}
\end{minipage} \hfill
\caption{Type-I wave: reflection and transmission rates for a finitely thick slab. We assumed $\omega\delta_0/c = 2.4 \times 10^{-2}$, cf. eq.~\eqref{eq_def_gamma}, and exaggerated the size of the photon mass for illustration purposes. Note that $R(\eta) > R(0)$ and $T(\eta) < T(0)$. In grey we show the approximation to the transmission rate as given by eq.~\eqref{eq_T_approx}, with $\eta = 0$. }
\label{fig_R_T_slab_type_I}
\end{figure}

In Sec.~\ref{sec_type_II_long_rate} we kept only terms up to second order in $\omega\delta_0/c$ in the reflection and transmission rates. Our main interest is in the transmission rate, cf. eqs.~\eqref{eq_z2} and~\eqref{eq_z3}, which may approximated at this level by
\begin{equation} \label{eq_T_approx}
\frac{|z_3|^2}{|z_2|^2} \approx 8\lambda \left( \frac{\omega\delta_0}{c} \right)^2 \left(1 - \eta^2 \right)  \, ,
\end{equation}
showing that $T(\eta) < T(0)$ as in the case of a semi-infinite medium, cf. Sec.~\ref{sec_type_I_rate}. In Fig.~\ref{fig_R_T_slab_type_I} we display the rates, cf. eq.~\eqref{eq_rates_slab_type_I}, as functions of $\overline{\xi} = \mathcal{D}/\delta_0$ keeping all orders in the small parameter $\omega\delta_0/c$. In grey we have the transmission rate (with $\eta = 0$) as given by eq.~\eqref{eq_T_approx}, showing that the approximation is excellent for slabs thicker than a few skin depths. The typical decline (saturation) of the transmission (reflection) rate is visible. The effects of a finite photon mass are minute, even for quite large (and unrealistic) values. From Fig.~\ref{fig_R_T_slab_type_I} we see that $R(\eta) > R(0)$ and, most importantly, we confirm that $T(\eta) < T(0)$, also if we do not use the approximate result above.

As a final remark we may check two opposing limits: no intermediate medium ($\mathcal{D} \rightarrow 0$) and a semi-infinite medium ($\mathcal{D} \rightarrow \infty$). In the former, there is no change in the propagation and we find that $T = 1$ and $R = 0$, as it should. In the latter limit, we find no transmission ``through" the slab, {\it i.e.}, $T = 0$, since it is now semi-infinite. As for the reflection rate, it reduces to the expected result, cf. eq.~\eqref{eq_R_rate_I}. Given that there is no transmission, the fact that $R \neq 1$ indicates that energy is dissipated within the medium -- we only have $R = 1$ in the limit of a perfect conductor.

\subsection{Mixed-type polarizations} \label{sec_type_II_long_slab}
\indent

Let us now analyse the richer case of mixed-type waves. The boundary conditions at $z = 0$ give us the following relations from the continuity of $\phi$ and ${\bf A}$: 
\begin{subequations}
\begin{eqnarray}
A_{\rm 0i}^{\rm (L)} + A_{\rm 0r}^{\rm (L)} & = & \frac{k'_{\rm L}}{k} \left( A_{\rm 0+}^{\rm (L)} + A_{\rm 0-}^{\rm (L)} \right)   \, , \label{eq_mixed_1}  \\
A_{\rm 0i}^{\rm (II)} - A_{\rm 0r}^{\rm (II)}   & = &  A_{\rm 0+}^{\rm (II)} - A_{\rm 0-}^{\rm (II)}  \, ,  \label{eq_mixed_2}  \\
A_{\rm 0i}^{\rm (L)} - A_{\rm 0r}^{\rm (L)}   & = &  A_{\rm 0+}^{\rm (L)} - A_{\rm 0-}^{\rm (L)}    \label{eq_mixed_3}  \, .
\end{eqnarray}
\end{subequations}
Furthermore, the continuity of ${\bf H}_\parallel$ gives
\begin{equation}\label{eq_mixed_4}
A_{\rm 0i}^{\rm (II)} + A_{\rm 0r}^{\rm (II)} = \frac{k'_{\rm T}}{k} \left( A_{\rm 0+}^{\rm (II)} + A_{\rm 0-}^{\rm (II)} \right)  \, , 
\end{equation}
while the information from the continuity of ${\bf B}_\perp$ is trivial. The continuity of ${\bf E}_\parallel$ does not deliver new constraints, since it merely reproduces eq.~\eqref{eq_mixed_2}. Here we have used that, in vacuum, both polarization modes have the same dispersion relation: $k_{\rm i}^{\rm L,T} = k_{\rm r}^{\rm L,T} = k_{\rm t}^{\rm L,T} = k$. Moreover, within the slab we have $k_{\rm +}^{\rm L} = k_{\rm -}^{\rm L} = k'_{\rm L}$ and $k_{\rm +}^{\rm T} = k_{\rm -}^{\rm T} = k'_{\rm T}$. Applying the boundary conditions to the second interface at $z = \mathcal{D}$ we obtain the following set of equations:
\begin{subequations}
\begin{eqnarray}
A_{\rm 0+}^{\rm (L)} e^{i\xi_{\rm L}} + A_{\rm 0-}^{\rm (L)} e^{-i\xi_{\rm L}} & = & \frac{k}{k'_{\rm L}} A_{\rm 0t}^{\rm (L)}  \label{eq_mixed_5} \, , \\
A_{\rm 0+}^{\rm (II)} e^{i\xi_{\rm T}} - A_{\rm 0-}^{\rm (II)} e^{-i\xi_{\rm T}}   & = &  A_{\rm 0t}^{\rm (II)}  \, , \label{eq_mixed_6}  \\
A_{\rm 0+}^{\rm (L)} e^{i\xi_{\rm L}} - A_{\rm 0-}^{\rm (L)} e^{-i\xi_{\rm L}}   & = &  A_{\rm 0t}^{\rm (L)} \, ,  \label{eq_mixed_7}  \\
A_{\rm 0+}^{\rm (II)} e^{i\xi_{\rm T}}  + A_{\rm 0-}^{\rm (II)} e^{-i\xi_{\rm T}}  & =  & \frac{k}{k'_{\rm T}} A_{\rm 0t}^{\rm (II)}  \label{eq_mixed_8}  \, .
\end{eqnarray}
\end{subequations}

In the expressions above we have $\xi_{\rm T} = (1+i)\overline{\xi}$, cf. eq.~\eqref{eq_def_xi}, and $\xi_{\rm L} = k'_{\rm L} \mathcal{D}$. The latter may be expanded to give (cf. eqs.~\eqref{eq_DRs} and~\eqref{eq_n})
\begin{equation}
\xi_{\rm L} = \frac{\mathcal{D}\omega}{c} \left[ 1 + \frac{i}{4}\left( \frac{\omega\delta_0}{c} \right)^2 \eta^2 \right] \, ,
\end{equation}
which contains a small, though finite, imaginary term. The real part of $\xi_{\rm L}$ will produce an oscillatory contribution independent of the photon mass, whereas the imaginary piece will generate a suppression term, cf. eq.~\eqref{eq_def_lambda}. Incidentally, we may write ${\rm Im}({\xi_{\rm L}}) = \mathcal{D}/\delta_{\rm L}$ with
\begin{equation} \label{eq_delta_L}
\delta_{\rm L} = \frac{4c}{\omega} \left( \frac{c}{\omega\delta_0} \right)^2 \frac{1}{\eta^2} 
\end{equation}
being the skin depth of the longitudinal mode, an impressively large number starkly contrasting to the sub-$\mu$m skin depth of its transverse counterparts.

We are not interested in the amplitudes within the slab, only in the reflected and transmitted amplitudes. With $k'_{\rm T}/k = \beta$, cf. eqs.~\eqref{eq_ratio} and~\eqref{eq_def_beta_bar}, and $k'_{\rm L}/k = \alpha$, cf. eq.~\eqref{eq_ratio_L}, the solution to the system of equations is
\begin{subequations}
\begin{eqnarray}
\frac{A_{\rm 0r}^{\rm (L)}}{A_{\rm 0i}^{\rm (L)} } & = & \frac{\left(  1 - \alpha^2 \right) \left( 1 - e^{2i\xi_{\rm L}} \right)  }{ e^{2i\xi_{\rm L}} (1 - \alpha)^2 - (1 + \alpha)^2 }  \label{eq_mix_slab_r_L} \, , \\
\frac{A_{\rm 0r}^{\rm (II)}}{A_{\rm 0i}^{\rm (II)} } &=& \frac{\left(  1 - \beta^2 \right) \left( 1 - e^{2i\xi_{\rm T}} \right) }{ e^{2i\xi_{\rm T}} (1 - \beta)^2 - (1 + \beta)^2 }  \label{eq_mix_slab_r_T} \, , \\
\frac{A_{\rm 0t}^{\rm (L)}}{A_{\rm 0i}^{\rm (L)} } & = & -\frac{4\alpha e^{i\xi_{\rm L}} }{ e^{2i\xi_{\rm L}} (1 - \alpha)^2 - (1 + \alpha)^2 }  \label{eq_mix_slab_t_L} \, ,
\\
\frac{A_{\rm 0t}^{\rm (II)}}{A_{\rm 0i}^{\rm (II)} } &=& -\frac{4\beta e^{i\xi_{\rm T}} }{ e^{2i\xi_{\rm T}} (1 - \beta)^2 - (1 + \beta)^2 } \label{eq_mix_slab_t_T}  \; .
\end{eqnarray}
\end{subequations}
These ratios have the same structure as those found for type-I waves, cf. eqs.~\eqref{eq_ratio_r_I_slab} and~\eqref{eq_ratio_t_I_slab}.

The squared amplitudes of the type-II component are given by the familiar expressions
\begin{equation}  \label{eq_squares_II}
\Bigg| \frac{A_{\rm 0r}^{\rm (II)}}{A_{\rm 0i}^{\rm (II)} } \Bigg|^2 = \frac{ |z_1|^2 }{ |z_2|^2 }   \quad {\rm and}  \quad  \Bigg| \frac{A_{\rm 0t}^{\rm (II)}}{A_{\rm 0i}^{\rm (II)} } \Bigg|^2 = \frac{ |z_3|^2 }{ |z_2|^2 }  \, ,
\end{equation}
with $z_1, z_2, z_3$ from eqs.~\eqref{eq_z1}-\eqref{eq_z3}. Moreover, keeping only the leading-order contributions in $\eta^2$, we find 
\begin{equation}\label{eq_squares_L}  
\Bigg| \frac{A_{\rm 0r}^{\rm (L)}}{A_{\rm 0i}^{\rm (L)} } \Bigg|^2  \approx  \frac{\sin^2\left( \frac{\mathcal{D}\omega}{c} \right) }{4} \frac{\eta^4}{1 + \eta^2}     \quad {\rm and}  \quad  \Bigg| \frac{A_{\rm 0t}^{\rm (L)}}{A_{\rm 0i}^{\rm (L)} } \Bigg|^2   \approx 1 \, ,
\end{equation}
where we used that $\exp(-2\mathcal{D}/\delta_{\rm L}) \approx 1$, cf. eq.~\eqref{eq_delta_L}. Neglecting terms of order $\mathcal{O}( |\rho|^4\eta^4)$ or higher, we finally obtain
\begin{subequations}
\begin{eqnarray}
R & \approx & \frac{ |z_1|^2 }{ |z_2|^2 } - |\rho|^2\eta^2 \left( \frac{ |z_1|^2 }{ |z_2|^2 } \right)_{\eta = 0} \label{eq_R_mixed_slab_final} \, , \\
T & \approx & \frac{ |z_3|^2 }{ |z_2|^2 } - |\rho|^2\eta^2 \left( \frac{ |z_3|^2 }{ |z_2|^2 } \right)_{\eta = 0} + |\rho|^2\eta^2 \label{eq_T_mixed_slab_final}  \, .
\end{eqnarray}
\end{subequations}

\begin{figure}[t!]
\begin{minipage}[b]{1.0\linewidth}
\includegraphics[width=\textwidth]{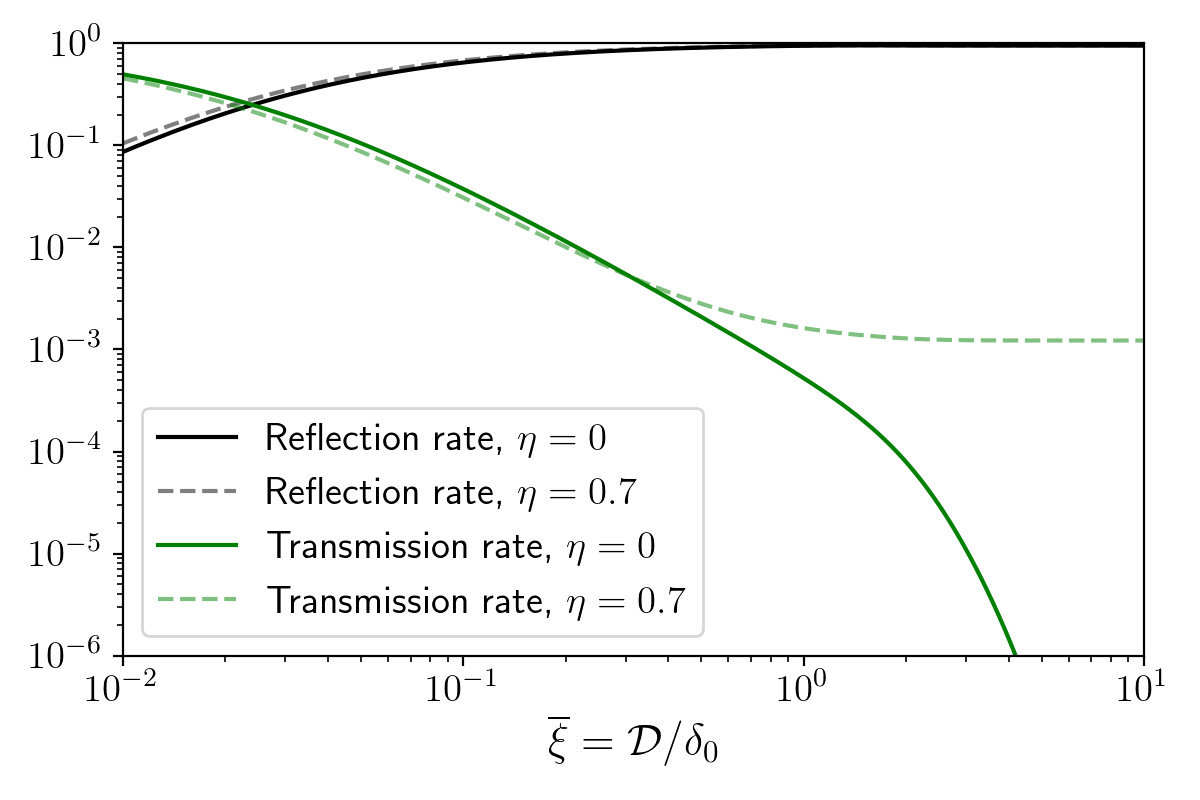}
\end{minipage} \hfill
\caption{Mixed-type wave: reflection and transmission rates for a finitely thick slab. We used $\omega\delta_0/c = 2.4 \times 10^{-2}$, cf. eq.~\eqref{eq_def_gamma}, and $|\rho| = 0.05$. The magnitude of the photon mass is exaggerated for illustration purposes. Note the change of sign of $T(0) - T(\eta)$ near $\overline{\xi} = \mathcal{D}/\delta_0 \approx 0.3$.}
\label{fig_R_T_slab_mixed_type}
\end{figure}

These rates are displayed in Fig.~\ref{fig_R_T_slab_mixed_type} where, as in Sec.~\ref{sec_type_I_slab}, we kept all orders in the small parameter $\omega\delta_0/c$. The saturation caused by the last term in eq.~\eqref{eq_T_mixed_slab_final} is visible -- it is expected as longitudinal photons always pass through, maintaining a minimal flux irrespective of the length traversed. Furthermore, we remark that $T(\eta) < T(0)$ only if $8\lambda \left( \omega\delta_0/c \right)^2 \gtrsim |\rho|^2$, as can be seen by plugging eq.~\eqref{eq_T_approx} into eq.~\eqref{eq_T_mixed_slab_final}. In fact, using the parameters from Fig.~\ref{fig_R_T_slab_mixed_type} we see that the sign change occurs at $\overline{\xi} = \mathcal{D}/\delta_0 \approx 0.3$, as visible in the figure.

\begin{figure}[t!]
\begin{minipage}[b]{1.0\linewidth}
\includegraphics[width=\textwidth]{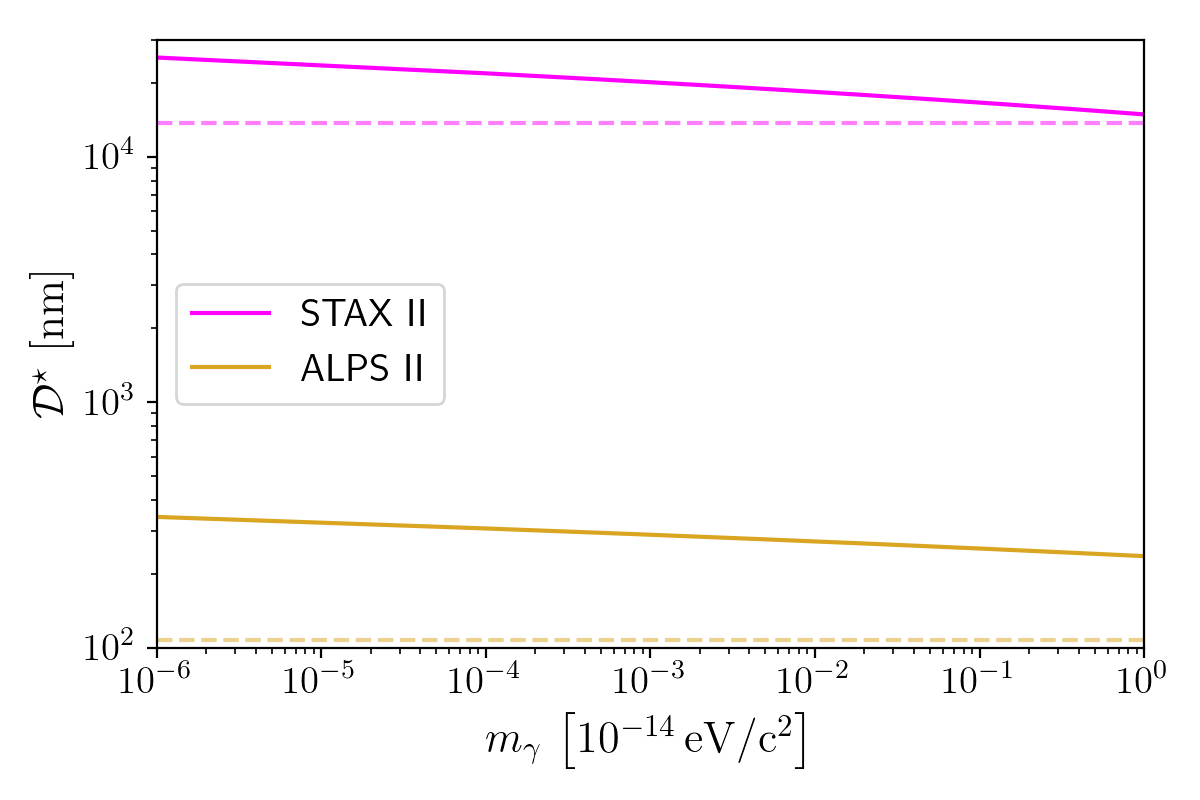}
\end{minipage} \hfill
\caption{Slab thickness beyond which the longitudinal component dominates the transmission rate, cf. eq.~\eqref{eq_D_star}. Here we used the parameters quoted in Tables~\ref{table_params} and~\ref{table_params_phys}. The dashed lines represent the threshold thickness where the two components of the background are equal, cf. eq.~\eqref{eq_bg}.}
\label{fig_D_star}
\end{figure}

We conclude that a conducting slab effectively acts as a filter that reflects and absorbs transverse modes, letting longitudinal waves pass and leading to the saturation of the dashed green curve shown in Fig.~\ref{fig_R_T_slab_mixed_type}. We thus expect that, behind a thick enough slab, only longitudinal waves will be left. The thickness $\mathcal{D}^\star$ beyond which the last term of eq.~\eqref{eq_T_mixed_slab_final} dominates is 
\begin{equation}  \label{eq_D_star}
\mathcal{D}^\star \approx -\frac{\delta_0}{2} \ln \left[ \frac{1}{8} \left( \frac{c}{\omega\delta_0} \right)^2 |\rho|^2 \eta^2 \right] \, ,
\end{equation}
which is shown in Fig.~\ref{fig_D_star} with $|\rho| = \eta$, cf. Sec~\ref{sec_exp_sens}. Like in Maxwell's theory, de Broglie-Proca waves with longer wavelengths need thicker slabs to have its transverse components effectively blocked. Incidentally, for the parameters assumed in Fig.~\ref{fig_R_T_slab_mixed_type} we find $\mathcal{D}^\star \approx 2.5$~nm, or $\overline{\xi} \approx 0.7$, which marks the onset of the saturation displayed by the dashed green curve.

At last, but not least, let us again look into the limiting behavior of the rates. For $\mathcal{D} \rightarrow 0$ we have $|z_1|^2 = 0$, so $R = 0$; similarly, $T = 1 + \mathcal{O}(\eta^4)$, as expected. Equations~\eqref{eq_R_mixed_slab_final} and~\eqref{eq_T_mixed_slab_final}, however, are not applicable at the limit $\mathcal{D} \rightarrow \infty$. In fact, $T = | \rho |^2\eta^2 > 0$. This is actually an artefact, since we have $\exp(-\mathcal{D}/\delta_{\rm L}) \approx 1$ for any finite (and physically meaningful) slab thickness.


\section{Experimental sensitivity}  \label{sec_exp_sens}
\indent

We have calculated the reflection and transmission rates for de Broglie-Proca waves propagating through a slab of conducting material. For the purely transverse type-I modes we found that the rates are essentially the same as in Maxwell's theory, particularly since $\eta \ll 1$, cf. Figs.~\ref{fig_R_T_slab_type_I}. For mixed-type waves the situation is different and the most important modification is the presence of the last term in eq.~\eqref{eq_T_mixed_slab_final}. Albeit potentially very small, it is barely attenuated, being therefore insensitive to the thickness of the slab.

In our context, signal photons are those whose flux (measured in photons per second) are compatible with the calculations of Sec.~\ref{sec_slab} with $\eta \neq 0$; these represent ``new physics". Background photons, on the other hand, are associated with no ``new physics" and are divided here in two main categories. The first is that of massless photons whose flux is controlled by the transmission rates with $\eta = 0$, {\it i.e.}, the standard Maxwell theory. We are therefore trying to disciminate the two cases based on the number of photon counts by a detector during a certain time interval. There are, however, other sources of radiation that would provide for extra photon counts, in particular blackbody emissions. Cooling the system to cryogenic temperatures $\lesssim \mathcal{O}({\rm 100 \, mK})$ helps mitigate the problem, though remaining thermal photons would still add to the dark count rate\footnote{The dark count rate measures the rate of photon recordings when no signal is expected. It is influenced by the properties of the detector, as well as by the environment.}, the second contribution to the background.

\begin{table}[t!]
\begin{tabular}{|c|c|c|c|c|}
\hline
 & Power & Photon energy & $\lambda_{\rm source}$ & $n_{\rm dc}$  \\ \hline
ALPS II & 30~W & 1.2~eV & 1064~nm & $10^{-6} \, \gamma$/s   \bigstrut[t] \\ \hline
STAX II & 1~MW & 124~$\mu$eV & 10~mm & $10^{-9} \, \gamma$/s   \bigstrut[t] \\ \hline
\end{tabular}
\caption{Key parameters of LSW experiments. Here we take the ALPS II setup~\cite{ALPS3, ALPS4} as a typical laser-based implementation, whereas the STAX II proposal is based on high-power microwave sources~\cite{STAX, STAX2, STAX3, ppt_STAX, STAX4}. These experiments include production and regeneration cavities to increase the effective power; the numbers quoted here do not include such refinements. The quantum efficiency of TESs has been demonstrated to reach $\gtrsim 95\%$ for near-infrared photons~\cite{TES_general, NIST_TES}, with a similar figure for microwave signals~\cite{ppt_STAX}, and is assumed to be maximal in this work. }  
\label{table_params}
\end{table}

Typical LSW setups deal with very low signal rates due to the regeneration probability $\lesssim 10^{-20}$~\cite{Ringwald}, thus requiring single-photon detectors. For example, a 2-W laser at 1000~nm provides $\sim 10^{19} \, \gamma/{\rm s}$ with an expected photon flux behind the wall of $\lesssim 0.1 \, \gamma/{\rm s}$. A very promising option for single-photon detection are transition-edge sensors (TES), whose dark count rate can be as low as $n_{\rm dc} \sim 10^{-6} \, \gamma/{\rm s}$ for near-infrared photons, as assumed for the ALPS II experiment~\cite{ALPS2, ALPS3, ALPS4, TES_general, dc_ALPS}. For the STAX proposal~\cite{STAX, STAX2, STAX3, ppt_STAX, STAX4} it is estimated that it can be lowered to $n_{\rm dc} \sim 10^{-9} \, \gamma/{\rm s}$ in its second stage (STAX II) by using a dedicated microwave-sensitive TES~\cite{ppt_STAX}. In what follows we neglect noise from other sources ({\it e.g.}, cosmic rays, ambient radiation, etc) and take the aforementioned dark count rates as the main detector-related contribution to the background. For the sake of concreteness,  in Table~\ref{table_params} we list benchmark values for latest LSW proposals based on lasers (ALPS II) and microwaves (STAX II).

Assuming that the beam emitted by the source keeps a constant profile throughout its passage within the slab, the number of transmitted photons is $N_{\rm t} = T(\eta) N_{\rm i}$. The number of photons of energy $\hbar\omega$ produced at the source during a time $\tau$ is $N_{\rm i} = \left( P_{\rm i}/\hbar\omega \right) \tau$, so that we have $N_{\rm t} = T(\eta) \left( P_{\rm i}/\hbar\omega \right) \tau$. The count rate, $n_{\rm t} = N_{\rm t}/\tau$, can be expressed as
\begin{equation} \label{eq_n_photons}
n_{\rm t}(\eta) = 5.0 \times 10^{18} \, {\rm Hz} \, \cdot T(\eta) \times \left( \frac{P_{\rm i}}{1 \, {\rm W}} \right) \left( \frac{\lambda_{\rm source}}{1000 \, {\rm nm}} \right)  \, ,
\end{equation}
and the numbers of signal and background photons are respectively
\begin{subequations}
\begin{eqnarray}
N_{\rm s} & = & T(\eta) \left( \frac{P_{\rm i}}{\hbar\omega} \right) \tau \label{eq_signal} \, , \\
N_{\rm b} & = & T(0) \left( \frac{P_{\rm i}}{\hbar\omega} \right) \tau + n_{\rm dc} \tau \label{eq_bg}  \, .
\end{eqnarray}
\end{subequations}

The expected transmission rates are very low, so any LSW experiment will have to detect single photons. We are thus effectively conducting a counting experiment with an expected number of signal photons $N_{\rm s} = n_{\rm s} \tau$ and background photons $N_{\rm b} = n_{\rm b} \tau$. The number of signal photons is reduced by two factors. The first is the so-called quantum efficiency, which is determined assuming that the signal consists of usual massless photons. Both teams report nearly maximal expected efficiencies for their respective TESs. Our signal, however, is composed of longitudinal photons, whose detection is not the design target of current or future detection devices. Therefore, the number of signal photons is $N_{\rm s} \rightarrow d_{\rm eff} N_{\rm s}$, where $d_{\rm eff} \leq 1$ models the efficiency of the detector device in measuring longitudinal photons. Following Refs.~\cite{Arias,stat1,stat2}, the statistical significance is  
\begin{equation} \label{eq_S12}
S_{12} = 2\left( \sqrt{N_{\rm b} + d_{\rm eff} N_{\rm s}} - \sqrt{N_{\rm b}} \right) \, .
\end{equation}
For a $95\%$~CL limit we use $S_{12} < 2$.

\begin{table}[t!]
\begin{tabular}{|c|c|c|c|}
\hline
 & $\eta$ & $\delta_0$ & $\omega\delta_0/c$  \\ \hline
ALPS II & $8.6 \times 10^{-15}$ & 3.9~nm &  $2.3 \times 10^{-2}$   \bigstrut[t] \\ \hline
STAX II & $8.1 \times 10^{-11}$ & 380~nm & $2.4 \times 10^{-4}$    \bigstrut[t] \\ \hline
\end{tabular}
\caption{Basic parameters evaluated using the benchmark values for the ALPS II~\cite{ALPS2, ALPS3, ALPS4} and STAX II setups~\cite{STAX, STAX2, STAX3, ppt_STAX, STAX4} as given in Table~\ref{table_params}. Here we take the conductivity of copper $\sigma = 5.9 \times 10^7 \, {\rm S/m}$ and a photon mass $m_\gamma = 10^{-14} \, {\rm eV/c^2}$ as references. The scalings for different values may be obtained by using eqs.~\eqref{eq_def_skin_numeric}, \eqref{eq_def_gamma} and~\eqref{eq_def_eta_numeric}.}\label{table_params_phys}
\end{table}

From now on we consider the parameters in Table~\ref{table_params}, as well as the calculated values summarized in Table~\ref{table_params_phys}. Using eq.~\eqref{eq_T_approx} with $\eta = 0$ we may compare the two contributions to the background, cf. eq.~\eqref{eq_bg}. For the laser source from ALPS II we find that the count rates are equal for $\mathcal{D} \approx 107$~nm, whereas for STAX II we have $\mathcal{D} \approx 14 \, \mu$m, cf. Fig.~\ref{fig_D_star}. For slabs thicker than these values, the dark count rate dominates and $N_{\rm b} = n_{\rm dc} \tau$. In what follows we will be working under this scenario.

In Sec.~\ref{sec_type_I_slab} we found that $T(\eta) < T(0)$ for type-I waves, so the signal $N_{\rm s}$ is smaller than the first term in $N_{\rm b}$, cf. eq.~\eqref{eq_bg}, which is itself smaller than the dark count rate, {\it i.e.}, $N_{\rm s} \ll N_{\rm b}$ for type-I waves. For mixed-type waves we found that the thickness-independent term in the transmission rate will be dominant for any thickness $\mathcal{D} \gtrsim \mathcal{D}^\star$, cf. eq.~\eqref{eq_D_star}, so $N_{\rm s} \sim |\rho|^2 \eta^2$, which is extremely small. We may then safely assume $N_{\rm s} \ll N_{\rm b}$ also in this case. The signal count rate will be then $n_{\rm s} < \left(S_{12}/d_{\rm eff}\right) \sqrt{ n_{\rm dc}/\tau}$, or more explicitly
\begin{eqnarray} \label{eq_lim_signal_final}
n_{\rm s} & < & 6.8 \times 10^{-6} \, {\rm Hz} \, \left( \frac{S_{12}}{2} \right) \left( \frac{1}{d_{\rm eff}} \right) \nonumber \\
&  & \quad \times \left( \frac{n_{\rm dc}}{10^{-6} \, {\rm Hz}} \right)^{1/2} \left( \frac{1 \, {\rm day}}{\tau} \right)^{1/2} \, .
\end{eqnarray}

From eq.~\eqref{eq_T_approx} we see that the limit for type-I waves scales with $\lambda^{-1} = \exp{\left(2\mathcal{D}/\delta_0\right)}$, worsening dramatically fast for thicker slabs. The possibility of working instead with thinner barriers is attractive, but it is nonetheless problematic. Most critical is the heating of the barrier due to the strong incident radiation, which would produce potentially overwhelming noise in the form of backbody radiation, as well as pose integrity risks for the thin metal sheet. Another issue might be the difficulty to shield the detector region behind the barrier against stray light. Moreover, the fabrication of nm-thick metal sheets is not trivial, as these must be homogeneous and sufficiently strong to withstand pressure gradients, since the apparatus is assumed to be in vacuum. Overall, the task of constraining the photon mass using type-I waves in a minimally feasible LSW setup seems hopeless.

Let us now move on to the more interesting case of mixed-type waves. Here the transmission rate has essentially two parts: one stemming from the transverse type-II component, scaling with $\lambda$, and one independent of the slab thickness originating from the longitudinal mode. This last feature means that the aforementioned problems related to the measurement of type-I waves are much less relevant here. In fact, the wall may be as thick as necessary to avoid issues with heating of the detector region, also facilitating the shielding from unwanted light from the source. The transverse part of the transmission rate will not contribute for any macroscopic thickness, leaving us with $T(\eta) \approx |\rho|^2\eta^2$, cf. eq.~\eqref{eq_T_mixed_slab_final}.

So far we have not presented an estimate for the magnitude of $\rho$, leaving it as an unknown parameter. In reality, it is fixed by the composition of the incident light, meaning that $\rho$ is determined by the emission mechanism in the source, where atoms undergo transitions between certain energy states. As pointed out by Goldhaber and Nieto~\cite{Nieto}, quantum mechanical transition amplitudes are of the form $T_{\rm fi} \sim \epsilon_\alpha \langle f | J^\alpha | i \rangle$, where $J^\alpha$ is a conserved 4-current and $\epsilon_\alpha$ is the polarization 4-vector of the photon. For longitudinal photons we have $\epsilon_\alpha \sim ( k_{\rm L}/\mu_\gamma, 0, 0, \omega/\mu_\gamma c)$, whereas $\epsilon_\alpha \sim (0, 1, 0, 0)$ for transverse ones polarized in the $x$ direction. This leads to $T^{\rm L}_{\rm fi} \sim \left( \mu_\gamma c/\omega \right) \langle f | J^{\rm z} | i \rangle$ and $T^{\rm T}_{\rm fi} \sim \langle f | J^{\rm x} | i \rangle$, so that 
\begin{equation} \label{eq_ratio_mq_amps}
\frac{ | T^{\rm L}_{\rm fi} |^2 }{ | T^{\rm T}_{\rm fi} |^2 } \sim  \eta^2 \, .
\end{equation}

The number of photons emitted by the atoms in the source is proportional to the emission cross section, $N \sim \sigma_{\rm em}$, which is itself proportional to the squared transition amplitude. From eq.~\eqref{eq_ratio_mq_amps} we have $N_{\rm L}/N_{\rm T} \sim \eta^2$, where $N_{\rm L}$ ($N_{\rm T}$) denotes the number of longitudinal (transverse) photons emitted. Finally, the amplitude of the electric field ($\sim A_{\rm 0i}$) is proportional to $\sqrt{N}$, implying that 
\begin{equation} \label{eq_rho_eta}
|\rho| \sim \eta \, .
\end{equation}
Therefore, we expect longitudinal photons to the extremely rare, unless the respective matrix elements are much larger than those for transverse photons. Incidentally, a similar argumentation may be used to estimate the detection efficiency for longitudinal photons, $d_{\rm eff}$. A real detector will only measure a fraction of the incident flux and the detection probability is determined by the absorption cross section, $\sigma_{\rm abs}$. For longitudinal photons we just argued that $\sigma_{\rm em} \sim \eta^2$, leading to eq.~\eqref{eq_rho_eta}, and we may naively expect that emission and absorption cross sections are similarly large, so $\sigma_{\rm abs} \sim \eta^2$ and consequently
\begin{equation} \label{eq_d_eff}
d_{\rm eff} \sim \eta^2 \, .
\end{equation}

Equations~\eqref{eq_rho_eta} and~\eqref{eq_d_eff} show that longitudinal photons are scarcely emitted and are even less likely to be detected with standard devices. These issues will have a strong impact on the experimental sensitivities: putting together eqs.~\eqref{eq_signal} and~\eqref{eq_lim_signal_final}, and using eqs.~\eqref{eq_rho_eta} and~\eqref{eq_d_eff}, we finally obtain
\begin{eqnarray} \label{eq_sens}
m_\gamma & < & 1.3 \times 10^{-4} \, {\rm eV/c^2} \, \left( \frac{S_{12}}{2} \right)^{1/6}  \left( \frac{1 \, {\rm day}}{\tau} \right)^{1/12} \\
&  \times  &  \left( \frac{n_{\rm dc}}{10^{-6} \, {\rm Hz}} \right)^{1/12}   \left( \frac{1 \, {\rm W}}{P_{\rm i}} \right)^{1/6} \left( \frac{1000 \, {\rm nm}}{\lambda_{\rm source}} \right)^{7/6}    \nonumber \, .
\end{eqnarray}
The result above assumes that the matrix elements are of similar magnitude, {\it i.e.}, $\langle f | J^{\rm x} | i \rangle \approx \langle f | J^{\rm z} | i \rangle$, as well as $|\rho| = \eta$ and $d_{\rm eff} = \eta^2$ (these are worst-case scenarios). We may now specialize eq.~\eqref{eq_sens} for the two benchmark setups considered here, cf. Table~\ref{table_params}. Assuming a total measurement time $\tau = 1$~year, we find
\begin{subequations}
\begin{eqnarray}
&  {\rm ALPS \,\, II}: \,\, & m_\gamma <  4.2 \times 10^{-5} \, {\rm eV/c^2}   \, , \label{eq_lim_ALPS} \\
&  {\rm STAX \,\, II}: \,\, & m_\gamma < 9.6 \times 10^{-11} \, {\rm eV/c^2}   \, ,    \label{eq_lim_STAX} 
\end{eqnarray}
\end{subequations}
both at $95\%$~CL.


\section{Concluding remarks}  \label{sec_conclusions}
\indent

In this paper we focused on the transmission of massive de Broglie-Proca waves from vacuum onto a nonpermeable, conducting material. We determined the amplitude ratios for oblique incidence onto a semi-infinite medium, subsequently specializing these results to normal incidence. Furthermore, we analysed the more interesting case of a slab of conducting material, also obtaining the associated reflection and transmission rates.

We showed that type-I waves satisfy the boundary conditions, but type-II and longitudinal modes separately do not (a linear combination of these solves the issue). By symmetry, there is no distinction between type-I and -II modes for normal incidence, thus indicating that it is the longitudinal mode that needs a transverse mode to support its passage through an interface. We also found that the contributions from type-I and -II transverse modes to the rates are identical for a semi-infinite medium or a slab. In fact, these are altogether not very different from those in Maxwell's theory.

Particularly interesting is the disparity between the skin depths for transverse and longitudinal modes. While the former behave similarly to their massless counterparts, having $\delta_{\rm T} \approx \delta_0$, the skin depth of the latter is practically infinite: longitudinal waves can barely distinguish a good conductor from vacuum. The longitudinal components are effectively unmodified by the passage through a conducting material medium. This, in turn, leads to the all-important term $\sim |\rho|^2 \eta^2$ in the transmission rates, which is independent of other system parameters, such as slab thickness, wave frequency or conductivity.

It is this longitudinal contribution that could provide a constraint of the photon mass. Unfortunately, the nature of the measurement and the low signal rate expected imply that data will be dominated by background noise. The very low dark count rates necessary for operating existing and future LSW experiments are a technological challenge, but even large improvements would have limited impact on the sensitivity due to the weak scaling displayed in eq.~\eqref{eq_sens}. Similarly, overextending the run time of the experiment would not be very effective. The best leverage is, in fact, the wavelength of the radiation. Scaling the sensitivity~\eqref{eq_lim_STAX}, we see that using 10-m radio waves (30~MHz) instead of 10-mm microwaves (keeping everything else fixed) would improve the sensitivity by a factor of $\sim 10^4$, reaching the ballpark of other limits based on terrestrial phenomena~\cite{Kroll2, Malta, PDG}.

In Sec.~\ref{sec_exp_sens} we estimated the magnitude of the proportion of longitudinal and transverse photons, finding that $|\rho| \sim \eta$. The exact numerical factor strongly depends on the microscopic details of the light source, in particular which atomic transitions are allowed for longitudinal photons and how fast these take place in comparison with those resulting in the emission of transverse photons. Incidentally, better understanding these issues would allow us to more precisely quantify the detector efficiency when measuring longitudinal photons, $d_{\rm eff}$, potentially helping optimize existing -- or design future -- devices. In this sense, the sensitivities obtained are pessimistic, since it is possible that the proportionality factors in eqs.~\eqref{eq_rho_eta} and~\eqref{eq_d_eff} weaken their dependence on $\eta$. Altogether, our final results, eqs.~\eqref{eq_lim_ALPS} and~\eqref{eq_lim_STAX} could be made more precise by means of a careful study of the quantum mechanical interaction of atoms with photons in the context of massive electrodynamics. These topics will be addressed elsewhere.

Finally, it is worth mentioning that a superconducting material could be used to construct the optical barrier. An immediate consequence of having a massive photon within a superconductor is a modification of the London penetration depth. In London's model the current density is ${\bf J} = -(e^2n/m_e c) {\bf A}$, where $e$ and $m_e$ are the charge and mass of the electron, and $n$ is the number density of the superconducting carriers~\cite{London}. Inserting this into eq.~\eqref{eq_ampere2} and ignoring the displacement current leads to the modified London equation
\begin{equation}
\nabla\times {\bf B} = - \left( \mu_\gamma^2 + \mu_0\frac{e^2 n}{m_e c} \right) {\bf A} \, ,
\end{equation}
{\it i.e.}, the London penetration depth is reduced if $\mu_\gamma \neq 0$ and the magnetic field is even more strongly expelled.

In order to study the consequences of shining (massive) light through a superconducting slab, it is necessary to analyse the subtleties of reflection and transmission at the interfaces, where adequate boundary conditions must be applied. This is a very interesting and non-trivial topic which deserves due attention. It lies beyond the scope of the present paper, but we shall pursue this investigation in a forthcoming paper.

In conclusion, the fact that effects of a finite photon mass are quadratic in $m_\gamma$ makes the study of massive electrodynamics quite challenging. Transverse components are severely attenuated over macroscopic distances and using thin barriers pose several difficulties. The possibility of measuring longitudinal modes would be the best chance to constrain $m_\gamma$, though impressive technological improvements in source power for long-wavelengths and corresponding single-photon detection capabilities would be required to reach limits competitive with other techniques based on terrestrial phenomena. Overall, our calculations support Goldhaber and Nieto's claim that the observation of longitudinal photons will be very difficult, if not impossible~\cite{Nieto}.


\begin{acknowledgments}
We thank the anonymous referees, J.A. Helay\"el-Neto and P. de Fabritiis for helpful comments, as well as P. Spagnolo for valuable information on the STAX project. PCM is indebted to Marina and Karoline Selbach for insightful discussions.
\end{acknowledgments}


\appendix

\end{document}